\documentclass[twocolumn,resetfootnote]{aastex701}
\def \champs{{\large \textsc{champs}}}
\newcolumntype{L}[1]{>{\raggedright\let\newline\\\arraybackslash\hspace{0pt}}m{#1}}
\newcolumntype{C}[1]{>{\centering\let\newline\\\arraybackslash\hspace{0pt}}m{#1}}
\newcolumntype{R}[1]{>{\raggedleft\let\newline\\\arraybackslash\hspace{0pt}}m{#1}}

\newcommand{\A}{\text{\normalfont\AA}}

\def \lya{Ly$\alpha$}

\def \h2{{\rm H_{2}}}

\def \Cii{[\ion{C}{2}]}
\def \oiii{[\ion{O}{3}]}

\def \dn4000{D_{{\rm n}}(4000) }

\def \x{$\times$}

\begin{document}

\title{CHAMPS: The COSMOS High-Redshift ALMA-MIRI Population Survey --\\A New Census of the Dusty Early Universe}

\suppressAffiliations

%%% FIRST AUTHOR
\author[0000-0002-9382-9832]{Andreas L. Faisst}
\affiliation{IPAC, California Institute of Technology, 1200 E. California Blvd. Pasadena, CA 91125, USA}
\email{afaisst@caltech.edu}
\correspondingauthor{Andreas L. Faisst}

%%% FIRST TIER %%%%
% co-PIs and core data reduction

% Felix Martinez
\author[0000-0002-9883-1413]{Felix Martinez III}
\email{fm5957@rit.edu}
\affiliation{Laboratory for Multiwavelength Astrophysics, School of Physics and Astronomy, Rochester Institute of Technology, 84 Lomb Memorial Drive, Rochester, NY 14623, USA}

% Jeyhan
\author[0000-0001-9187-3605]{Jeyhan S. Kartaltepe}
\affiliation{Laboratory for Multiwavelength Astrophysics, School of Physics and Astronomy, Rochester Institute of Technology, 84 Lomb Memorial Drive, Rochester, NY 14623, USA}
\email{jeyhan@astro.rit.edu}

%Manuel Aravena
\author[0000-0002-6290-3198]{Manuel Aravena}
\affiliation{Instituto de Estudios Astrofisicos, Facultad de Ingenieria y Ciencias, Universidad Diego Portales, Av. Ej\'ercito 441, Santiago 8370191, Chile}
\affiliation{Millenium Nucleus for Galaxies (MINGAL), Av. Ej\'ercito 441, Santiago 8370191, Chile}
\email{}

%Caitlin Casey,
\author[0000-0002-0930-6466]{Caitlin M. Casey}
\email{cmcasey@ucsb.edu}
\affiliation{Department of Physics, University of California, Santa Barbara, Santa Barbara, CA 93106, USA}
\affiliation{Cosmic Dawn Center (DAWN), Denmark} 
\email{}

%John Silverman,
\author[0000-0002-0000-6977]{John D. Silverman}
\affiliation{Kavli Institute for the Physics and Mathematics of the Universe (WPI), The University of Tokyo, Kashiwa, Chiba 277-8583, Japan}
\affiliation{Department of Astronomy, School of Science, The University of Tokyo, 7-3-1 Hongo, Bunkyo, Tokyo 113-0033, Japan}
\email{}

%Sune Toft
\author[0000-0003-3631-7176]{Sune Toft}
\affiliation{Cosmic Dawn Center (DAWN), Denmark} 
\affiliation{Niels Bohr Institute, University of Copenhagen, Jagtvej 128, DK-2200, Copenhagen, Denmark}
\email{sune@nbi.ku.dk}

% Ezequiel Treister
\author[0000-0001-7568-6412]{Ezequiel Treister}
\email{etreiste@astro.puc.cl}
\affiliation{
Instituto de Astrofísica, Facultad de F{\'i}sica, Pontificia Universidad Cat{\'o}lica de Chile, Casilla 306, Santiago 22, Chile}
\affiliation{Instituto de Alta Investigaci\'on, Universidad de Tarapac\'a, Casilla 7D, Arica, Chile}

%Jorge Zavala
\author[0000-0002-7051-1100]{Jorge A. Zavala}
\email{jzavala@umass.edu}
\affiliation{Department of Astronomy, University of Massachusetts Amherst, 710 N Pleasant Street, Amherst, MA 01003, USA}

%%% Second Tier %%%%%
% Others who have actively contributed but are not in first tier including

% Hiddo Algera
\author[0000-0002-4205-9567]{Hiddo S. B. Algera}
\email{hsbalgera@asiaa.sinica.edu.tw}
\affiliation{
Institute of Astronomy and Astrophysics, Academia Sinica, 11F of Astronomy-Mathematics Building, No.1, Sec. 4, Roosevelt Rd,
Taipei 106319, Taiwan, R.O.C.}

% Andrew Battisti
\author[0000-0003-4569-2285]{Andrew J. Battisti}
\affil{International Centre for Radio Astronomy Research, University of Western Australia, 35 Stirling Hwy, Crawley, WA 6009, Australia}
\affil{Research School of Astronomy and Astrophysics, Australian National University, Cotter Road, Weston Creek, ACT 2611, Australia}
\email{}

% Michele Catone
\author[0009-0001-8880-944X]{Michele Catone}
\affiliation{Dipartimento di Fisica e Astronomia Galileo Galilei Universit{\`a} degli Studi di Padova, Vicolo dell’Osservatorio 3, 35122 Padova Italy}
\affiliation{Istituto Nazionale di Astrofisica (INAF), Osservatorio Astronomico di Padova,
Vicolo dell’Osservatorio 5, 35122, Padova, Italy}
\email{michele.catone@studenti.unipd.it}

%Jaclyn Champagne
\author[0000-0002-6184-9097]{Jaclyn B. Champagne}
\affiliation{Space Telescope Science Institute, 3700 San Martin Drive, Baltimore, MD 21218, USA}
\email{jchampagne@stsci.edu}

% Yingjie Cheng
\author[0000-0001-8551-071X]{Yingjie Cheng}
\email{yingjiec@uw.edu}
\affiliation{Department of Astronomy, The University of Washington, Seattle, WA 98195, USA}
\affiliation{Department of Astronomy, University of Massachusetts Amherst, 710 N Pleasant Street, Amherst, MA 01003, USA}

% Rasha Samir
\author[0000-0003-2716-8332]{Rasha M. Samir}
\email{}
\affiliation{Department of Astronomy, National Research Institute of Astronomy and Geophysics (NRIAG), Cairo, 11421, Egypt}

% Max Franco
\author[0000-0002-3560-8599]{Maximilien Franco}
\email{maximilien.franco@cea.fr}
\affiliation{Universit{\'e} Paris-Saclay, Universit{\'e} Paris Cit{\'e}, CEA, CNRS, AIM, 91191 Gif-sur-Yvette, France}

% Kyle Finner
\author[0000-0002-4462-0709]{Kyle Finner}
\email{kfinner@ipac.caltech.edu}
\affiliation{IPAC, California Institute of Technology, 1200 E. California Blvd. Pasadena, CA 91125, USA}

% Carter Flayhart
\author[0009-0000-0272-5468]{Carter Flayhart}
\email{cbf2388@rit.edu}
\affiliation{Laboratory for Multiwavelength Astrophysics, School of Physics and Astronomy, Rochester Institute of Technology, 84 Lomb Memorial Drive, Rochester, NY 14623, USA}

% Fabrizio Gentile
\author[0000-0002-8008-9871]{Fabrizio Gentile}
\affiliation{Universit{\'e} Paris-Saclay, Universit{\'e} Paris Cit{\'e}, CEA, CNRS, AIM, 91191 Gif-sur-Yvette, France}
\affiliation{INAF-Osservatorio di Astrofisica e Scienza dello Spazio, Via Gobetti 93/3, 40129, Bologna, Italy}
\email{}

% Ghassem Gozaliasl
\author[0000-0002-0236-919X]{Ghassem Gozaliasl}
\email{ghassem.gozaliasl@gmail.com}
\affiliation{Department of Computer Science, Aalto University, P.O. Box 15400, FI-00076 Espoo, Finland}
\affiliation{Department of Physics, University of, P.O. Box 64, FI-00014 Helsinki, Finland}

% Santosh Harish
\author[0000-0003-0129-2079]{Santosh Harish}
\email{harish.santosh@gmail.com}
\affiliation{Space Telescope Science Institute, 3700 San Martin Drive, Baltimore, MD 21218, USA}

% Shuowen Jin
\author[0000-0002-8412-7951]{Shuowen Jin}
\email{shuowen.jin@gmail.com}
\affiliation{Cosmic Dawn Center (DAWN), Denmark}
\affiliation{DTU Space, Technical University of Denmark, Elektrovej 327, 2800 Kgs. Lyngby, Denmark}

% Yu-Heng Lin
\author[0000-0001-8792-3091]{Yu-Heng Lin}
\email{ianlin@ipac.caltech.edu}
\affiliation{IPAC, California Institute of Technology, 1200 E. California Blvd. Pasadena, CA 91125, USA}

% Lunjun Liu
\author[0009-0004-1270-2373]{Lun-Jun Liu}
\email{lliu@caltech.edu}
\affiliation{California Institute of Technology, 1200 E. California Blvd., Pasadena, CA, 91125 USA}

% Arianna Long
\author[0000-0002-7530-8857]{Arianna S. Long}
\affiliation{Department of Astronomy, The University of Washington, Seattle, WA 98195, USA}
\email{aslong@uw.edu}

% Crystal Martin
\author[0000-0001f-9189-7818]{Crystal L. Martin}
\email{}
\affiliation{Department of Physics, University of California, Santa Barbara, Santa Barbara, CA 93106, USA}

% Bahram Mobasher
\author[0000-0001-5846-4404]{Bahram Mobasher}
\affiliation{Department of Physics and Astronomy, University of California, Riverside, 900 University Avenue, Riverside, CA 92521, USA}
\email{}

% Francesca Pozzi
\author[0000-0002-7412-647X]{Francesca Pozzi}
\affiliation{University of Bologna – Department of Physics and Astronomy ``Augusto Righi'' (DIFA), Via Gobetti 93/2, 40129 Bologna, Italy}
\affiliation{INAF-Osservatorio di Astrofisica e Scienza dello Spazio, Via Gobetti 93/3, 40129, Bologna, Italy}
\email{}

% Huimin Qu
\author[0000-0002-4462-0709]{Huimin Qu}
\email{}
\affiliation{IPAC, California Institute of Technology, 1200 E. California Blvd. Pasadena, CA 91125, USA}

% Chuan Tian 
\author[0000-0003-4056-7071]{Chuan Tian}
\email{}
\affiliation{Department of Physics, Yale University, New Haven, CT 06520, USA}

% Mattia Vaccari
\author[0000-0002-6748-0577]{Mattia Vaccari}
\affiliation{Inter-University Institute for Data Intensive Astronomy (IDIA), Department of Astronomy, University of Cape Town, 7701 Rondebosch, Cape Town, South Africa}
\affiliation{Department of Physics and Astronomy, University of the Western Cape, 7535 Bellville, Cape Town, South Africa}
\affiliation{INAF-Osservatorio di Astrofisica e Scienza dello Spazio, Via Gobetti 93/3, 40129, Bologna, Italy}
\email{mattia.vaccari@gmail.com}

% Wuji Wang
\author[0000-0002-7964-6749]{Wuji Wang}
\affiliation{IPAC, California Institute of Technology, 1200 E. California Blvd. Pasadena, CA 91125, USA}
\email{wujiwang@ipac.caltech.edu}

% Can Xu
\author[0000-0003-3903-6935]{Can Xu} 
\email{}
\affiliation{School of Astronomy and Space Science, Nanjing University, Nanjing 210093, China}
\affiliation{Key Laboratory of Modern Astronomy and Astrophysics, Nanjing University, Nanjing 210093, China}
\affiliation{Kavli Institute for the Physics and Mathematics of the Universe (WPI), The University of Tokyo, Kashiwa, Chiba 277-8583, Japan}
\affiliation{Center for Data-Driven Discovery, Kavli IPMU (WPI), UTIAS, The University of Tokyo, Kashiwa, Chiba 277-8583, Japan}

%%% Third Tier %%%%
% co-Is on proposal and others.

%Hollis Akins
\author[0000-0003-3596-8794]{Hollis B. Akins}
\email{hollis.akins@gmail.com}
\altaffiliation{NSF Graduate Research Fellow}
\affiliation{The University of Texas at Austin, 2515 Speedway Blvd Stop C1400, Austin, TX 78712, USA}

%Micaela Bagley
\author[0000-0002-9921-9218]{Micaela B. Bagley}
\affiliation{The University of Texas at Austin, 2515 Speedway Blvd Stop C1400, Austin, TX 78712, USA}
\email{}

%Angela Bongiorno
\author[0000-0002-0101-6624]{Angela Bongiorno}
\email{angela.bongiorno@inaf.it}
\affiliation{ INAF-Observatory of Rome, via Frascati 33, 00074 Monteporzio Catone, Italy}

%Kevin Cooke
\author[0000-0002-2200-9845]{Kevin C. Cooke}
\affiliation{Association of Public and Land-grant Universities, 1220 L Street NW, Suite 1000, Washington, DC, US 20005}
\email{}

%Olivia Cooper
\author[0000-0003-3881-1397]{Olivia R. Cooper}
\email{ocooper@utexas.edu}
\affiliation{Department for Astrophysical \& Planetary Science, University of Colorado, Boulder, CO 80309, USA}

% Simon Hempel-Costello
\author[0009-0006-8917-119X]{Simon Hempel-Costello}
\email{shempelc@caltech.edu}
\affiliation{California Institute of Technology, 1200 E. California Blvd., Pasadena, CA, 91125 USA}

%Ivan Delvecchio
\author[0000-0001-8706-2252]{Ivan Delvecchio}
\email{ivan.delvecchio@inaf.it}
\affiliation{INAF-Osservatorio di Astrofisica e Scienza dello Spazio, Via Gobetti 93/3, 40129, Bologna, Italy}

%Nicole Drakos
\author[0000-0003-4761-2197]{Nicole E. Drakos}
\email{ndrakos@hawaii.edu}
\affiliation{Department of Physics and Astronomy, University of Hawaii, Hilo, 200 W Kawili St, Hilo, HI 96720, USA}

%Andrea Francesco Maria Enia
\author[0000-0002-4462-0709]{Andrea Enia}
\email{andreaenia@gmail.com}
\affiliation{INAF-Osservatorio di Astrofisica e Scienza dello Spazio, Via Gobetti 93/3, 40129, Bologna, Italy}

%Steven Finkelstein
\author[0000-0001-8519-1130]{Steven L. Finkelstein}
\affiliation{The University of Texas at Austin, 2515 Speedway Blvd Stop C1400, Austin, TX 78712, USA}
\email{stevenf@astro.as.utexas.edu}

%Seiji Fujimoto
\author[0000-0001-7201-5066]{Seiji Fujimoto}
\email{}
\affiliation{David A. Dunlap Department of Astronomy and Astrophysics, University of Toronto, 50 St. George Street, Toronto, Ontario, M5S 3H4, Canada}

%Steven Gillman
\author[0000-0001-9885-4589]{Steven Gillman}
\affiliation{Cosmic Dawn Center (DAWN), Denmark}
\affiliation{DTU Space, Technical University of Denmark, Elektrovej 327, 2800 Kgs. Lyngby, Denmark}
\email{srigi@space.dtu.dk}

%Carlos Gomez-Guijarro
\author[0000-0002-4085-9165]{Carlos Gomez-Guijarro}
\email{carlos.gomezguijarro@cea.fr}
\affiliation{Universit{\'e} Paris-Saclay, Universit{\'e} Paris Cit{\'e}, CEA, CNRS, AIM, 91191 Gif-sur-Yvette, France}

%Jorge González López
\author{Jorge Gonz\'alez-L\'opez}
\affiliation{Instituto de Astrof\'isica, Facultad de Física, Pontificia Universidad Cat\'olica de Chile, Santiago 7820436, Chile 3 Las Campanas}
\affiliation{Observatory, Carnegie Institution of Washington, Ra\'u Bitr\'an 1200, La Serena, Chile}
\email{jgonzalez@carnegiescience.edu}

%Michaela Hirschmann
\author[0000-0002-3301-3321]{Michaela Hirschmann}
\affiliation{Institute of Physics, GalSpec, Ecole Polytechnique Federale de Lausanne, Observatoire de Sauverny, Chemin Pegasi 51, 1290 Versoix, Switzerland}
\affiliation{INAF, Astronomical Observatory of Trieste, Via Tiepolo 11, 34131 Trieste, Italy}
\email{}

%Olivier Ilbert
\author[0000-0002-7303-4397]{Olivier Ilbert}
\email{olivier.ilbert@lam.fr}
\affiliation{Aix Marseille Univ, CNRS, CNES, LAM, Marseille, France}

%Kei Ito
\author[0000-0002-9453-0381]{Kei Ito}
\email{kei.ito@astron.s.u-tokyo.ac.jp}
\affiliation{Cosmic Dawn Center (DAWN), Denmark}
\affiliation{DTU Space, Technical University of Denmark, Elektrovej 327, 2800 Kgs. Lyngby, Denmark}

%Knud Jahnke
\author[0000-0003-3804-2137]{Knud Jahnke}
\affiliation{Max Planck Institute for Astronomy, K\"onigstuhl 17, D-69117 Heidelberg, Germany}
\email{}

%Boris Sindhu Kalita
\author[0000-0001-9215-7053]{Boris S. Kalita}
\email{boris.kalita@pku.edu.cn}
\affiliation{Kavli Institute for Astronomy and Astrophysics, Peking University, Beijing 100871, People{\textquotesingle}s Republic of China}
\affiliation{Kavli Institute for the Physics and Mathematics of the Universe (WPI), The University of Tokyo, Kashiwa, Chiba 277-8583, Japan}

%Daichi Kashino
\author[0000-0001-9044-1747]{Daichi Kashino}
\email{}
\affiliation{National Astronomical Observatory of Japan, 2-21-1 Osawa, Mitaka, Tokyo 181-8588, Japan}

%Vasily Kokorev
\author[0000-0002-5588-9156]{Vasily Kokorev}
\affiliation{The University of Texas at Austin, 2515 Speedway Blvd Stop C1400, Austin, TX 78712, USA}
\email{vasily.kokorev.astro@gmail.com}

% Thomas Lai
\author[0000-0001-8490-6632]{Thomas S.-Y. Lai}
\email{thomaslai.astro@gmail.com}
\affiliation{IPAC, California Institute of Technology, 1200 E. California Blvd. Pasadena, CA 91125, USA}

%Erini Lambrides
\author[0000-0003-3216-7190]{Erini Lambrides}\altaffiliation{NPP Fellow}
\email{}
\affiliation{NASA-Goddard Space Flight Center, Code 662, Greenbelt, MD, 20771, USA}

%Daizhong Liu
\author[0000-0001-9773-7479]{Daizhong Liu}
\email{dzliu@pmo.ac.cn}
\affiliation{Purple Mountain Observatory, Chinese Academy of Sciences, 10 Yuanhua Road, Nanjing 210023, China}

%Georgios Magdis
\author[0000-0002-4872-2294]{Georgios E. Magdis}
\email{geoma@space.dtu.dk}
\affiliation{Cosmic Dawn Center (DAWN), Denmark} 
\affiliation{DTU Space, Technical University of Denmark, Elektrovej 327, 2800 Kgs. Lyngby, Denmark}
\affiliation{Niels Bohr Institute, University of Copenhagen, Jagtvej 128, DK-2200, Copenhagen, Denmark}

%Vincenzo Mainieri
\author[0000-0002-1047-9583]{Vincenzo Mainieri}
\affiliation{European Southern Observatory, Karl-Schwarzschild-Straße 2, D-85748 Garching bei München, Germany}
\email{}

%Sinclaire Manning
\author[0000-0003-0415-0121]{Sinclaire M. Manning}
\email{smanning@astro.umass.edu}
\affiliation{Department of Astronomy, University of Massachusetts Amherst, 710 N Pleasant Street, Amherst, MA 01003, USA}

%Claudia Maraston
\author[0000-0001-7711-3677]{Claudia Maraston}
\email{}
\affiliation{Institute of Cosmology and Gravitation, University of Portsmouth, Dennis Sciama Building, Burnaby Road, Portsmouth, PO13FX, United Kingdom}

%Richard Massey
\author[0000-0002-6085-3780]{Richard Massey}
\affil{Department of Physics, Centre for Extragalactic Astronomy, Durham University, South Road, Durham DH1 3LE, UK}
\email{}

%Henry McCracken
\author[0000-0002-9489-7765]{Henry Joy McCracken}
\email{hjmcc@iap.fr}
\affiliation{Institut d’Astrophysique de Paris, UMR 7095, CNRS, and Sorbonne Université, 98 bis boulevard Arago, F-75014 Paris, France}

%Jed McKinney
\author[0000-0002-6149-8178]{Jed McKinney}
\affiliation{The University of Texas at Austin, 2515 Speedway Blvd Stop C1400, Austin, TX 78712, USA}
\affiliation{Cosmic Frontier Center, The University of Texas at Austin, 2515 Speedway Blvd Stop C1400, Austin, TX 78712, USA}
\email{}

%Wilfried Mercier
\author[0000-0001-6865-499X]{Wilfried Mercier}
\email{wilfried.mercier@lam.fr}
\affiliation{Aix Marseille Univ, CNRS, CNES, LAM, Marseille, France}

%Thibaud Moutard
\author[0000-0002-3305-9901]{Thibaud Moutard}
\email{thibaud.moutard@lilo.org}
\affiliation{European Space Agency (ESA), European Space Astronomy Centre (ESAC), Camino Bajo del Castillo s/n, 28692 Villanueva de la
Cañnda, Madrid, Spain}

%Masafusa Onoue
\author[0000-0003-2984-6803]{Masafusa Onoue}
\email{monoue@icloud.com}
\affiliation{Waseda Institute for Advanced Study (WIAS), Waseda University, Shinjuku, Tokyo 169-0051, Japan}

%Louise Paquereau
\author[0000-0003-2397-0360]{Louise Paquereau} 
\email{louise.paquereau@chalmers.se}
\affiliation{Department of Physics and Astronomy, Chalmers University of Technology, SE-412 96 Gothenburg, Sweden}

%Alvio Renzini
\author[0000-0002-7093-7355]{Alvio Renzini}
\affiliation{Istituto Nazionale di Astrofisica (INAF), Osservatorio Astronomico di Padova,
Vicolo dell’Osservatorio 5, 35122, Padova, Italy}
\email{}

%Jason Rhodes
\author[0000-0002-4485-8549]{Jason Rhodes}
\affiliation{Jet Propulsion Laboratory, California Institute of Technology, 4800 Oak Grove Drive, Pasadena, CA 91001, USA}
\email{}

%Robert Rich
\author[0000-0003-0427-8387]{R. Michael Rich}
\affiliation{Department of Physics and Astronomy, UCLA, PAB 430 Portola Plaza, Box 951547, Los Angeles, CA 90095-1547}
\email{}

%Brant Robertson
\author[0000-0002-4271-0364]{Brant E. Robertson}
\affiliation{Department of Astronomy and Astrophysics, University of California, Santa Cruz, 1156 High Street, Santa Cruz, CA 95064, USA}
\email{brant@ucsc.edu}

%Giulia Rodighiero
\author[0000-0002-9415-2296]{Giulia Rodighiero}
\affiliation{Dipartimento di Fisica e Astronomia Galileo Galilei Universit{\`a} degli Studi di Padova, Vicolo dell’Osservatorio 3, 35122 Padova Italy}
\affiliation{Istituto Nazionale di Astrofisica (INAF), Osservatorio Astronomico di Padova,
Vicolo dell’Osservatorio 5, 35122, Padova, Italy}
\email{giulia.rodighiero@unipd.it}

%David Sanders
\author[0000-0002-1233-9998]{David B. Sanders}
\affiliation{Institute for Astronomy, University of Hawai’i at Manoa, 2680 Woodlawn Drive, Honolulu, HI 96822, USA}
\email{}

%Claudia Scarlata
\author[0000-0002-9136-8876]{Claudia M. Scarlata}
\affiliation{Minnesota Institute for Astrophysics, School of Physics and Astronomy, University of Minnesota, 316 Church Street SE, Minneapolis, MN 55455, USA}
\email{}

%Nick Scoville
\author[0000-0002-0438-3323]{Nick Scoville}
\affiliation{California Institute of Technology, 1200 E. California Blvd., Pasadena, CA, 91125 USA}
\email{}

%Kartik Sheth
\author[0000-0002-5496-4118]{Kartik Sheth}
\email{astrokartik@gmail.com}
\affiliation{Aix Marseille Univ, CNRS, CNES, LAM, Marseille, France}

%Marko Shuntov
\author[0000-0002-7087-0701]{Marko Shuntov}
\email{marko.shuntov@nbi.ku.dk}
\affiliation{Cosmic Dawn Center (DAWN), Denmark} 
\affiliation{Niels Bohr Institute, University of Copenhagen, Jagtvej 128, DK-2200, Copenhagen, Denmark}
\affiliation{University of Geneva, 24 rue du G\'en\'eral-Dufour, 1211 Gen\`eve 4, Switzerland}

%Laura Sommovigo
\author[0000-0002-2906-2200]{Laura Sommovigo}
\email{laura.sommovigo.work@gmail.com}
\affiliation{Center for Computational Astrophysics, Flatiron Institute, 162 5th Avenue, New York, NY 10010, USA}

%Martin Sparre
\author[0000-0002-9735-3851]{Martin Sparre}
\affiliation{Institut f\"ur Physik und Astronomie, Universit\"at Potsdam, Karl-Liebknecht-Str.\,24/25, 14476 Golm, Germany}
\affiliation{Leibniz-Institut f\"ur Astrophysik Potsdam (AIP), An der Sternwarte 16, 14482 Potsdam, Germany}
\email{}

%Tomoko Suzuki
\author[0000-0002-3560-1346]{Tomoko L. Suzuki}
\affiliation{Kavli Institute for the Physics and Mathematics of the Universe (WPI), The University of Tokyo, Kashiwa, Chiba 277-8583, Japan}
\email{}

%Margherita Talia
\author[0000-0003-4352-2063]{Margherita Talia}
\email{margherita.talia2@unibo.it}
\affiliation{University of Bologna – Department of Physics and Astronomy ``Augusto Righi'' (DIFA), Via Gobetti 93/2, 40129 Bologna, Italy}
\affiliation{INAF-Osservatorio di Astrofisica e Scienza dello Spazio, Via Gobetti 93/3, 40129, Bologna, Italy}

%Benny Trakhtenbrot
\author[0000-0002-3683-7297]{Benny Trakhtenbrot}
\email{benny.trakht@gmail.com}
\affiliation{School of Physics and Astronomy, Tel Aviv University, Tel Aviv 69978, Israel}

%Francesco Valentino
\author[0000-0001-6477-4011]{Francesco Valentino}
\email{fmava@dtu.dk}
\affiliation{Cosmic Dawn Center (DAWN), Denmark} 
\affiliation{DTU Space, Technical University of Denmark, Elektrovej 327, 2800 Kgs. Lyngby, Denmark}

%Brittany Vanderhoof
\author[0000-0002-8163-0172]{Brittany N. Vanderhoof}
\email{bvanderhoof@stsci.edu}
\affiliation{Space Telescope Science Institute, 3700 San Martin Drive, Baltimore, MD 21218, USA}

%Eleni Vardoulaki
\author[0000-0002-4437-1773]{Eleni Vardoulaki}
\email{}
\affiliation{IAASARS/National Observatory Athens, Hill of Nymps, Athens 11810, Greece}

%Aswin Vijayan
\author[0000-0002-1905-4194]{Aswin P. Vijayan}
\email{}
\affiliation{Astronomy Centre, University of Sussex, Falmer, Brighton BN1 9QH, UK}

%Katherine Whitaker
\author[0000-0001-7160-3632]{Katherine E. Whitaker}
\affiliation{Department of Astronomy, University of Massachusetts Amherst, 710 N Pleasant Street, Amherst, MA 01003, USA}
\affiliation{Cosmic Dawn Center (DAWN), Denmark}
\email{kwhitaker@astro.umass.edu}

%Stephen Wilkins
\author[0000-0003-3903-6935]{Stephen M.~Wilkins} 
\affiliation{Astronomy Centre, University of Sussex, Falmer, Brighton BN1 9QH, UK}
\affiliation{Institute of Space Sciences and Astronomy, University of Malta, Msida MSD 2080, Malta}
\email{s.wilkins@sussex.ac.uk}

\collaboration{all}{(Affiliations can be found after the references)}

\begin{abstract}

Understanding how dust forms and evolves over cosmic time is a key open problem in extragalactic astrophysics.
Galaxies in the early Universe were once thought to be mostly dust-poor, but recent discoveries of heavily dust-obscured galaxies near the Epoch of Reionization challenge models of early dust production and raise new questions about the true cosmic star formation rate density.
In this paper, we present the {\em COSMOS High-Redshift ALMA-MIRI Population Survey} (\champs), which sheds light on the dusty early universe at $z>4$.
\champs~is a new 144h ALMA $1.2\,{\rm mm}$ (band 6) large program mapping the combined JWST/NIRCam$+$MIRI footprint in the COSMOS field. It covers $0.2\,{\rm deg^2}$ with a synthesized beam size of $1.1\arcsec$ and a sensitivity of $0.14\,{\rm mJy/beam}$. It is complemented, alongside JWST, by ancillary X-ray, UV, optical, sub-mm, and radio data.
\champs, optimized as a blind survey, detects above $5\sigma$ approximately $385$ individual dusty high-redshift sub-mm sources, $68$ dusty AGN, and a dozen dusty galaxies at $z>6$. The galaxies are largely not detected in JWST/F150W, making them prototypical near-IR dark galaxies.
We showcase three of the highest-redshift most dusty and massive candidate galaxies detected in \champs~with redshifts up to $z\approx7.5$.
\champs~also enables stacks of thousands of galaxies grouped in bins of various physical parameters. Its sub-mm coverage of dust-attenuated JWST-identified galaxies provides robust constraints on dust masses and obscured star formation rates to study early dust formation and evolution.
In addition, \champs~enables a blind search for luminous CO line emitters at $z=0.4$–$2.7$ and \Cii$_{\rm 158\mu m}$ emitters at $z\approx6.7$.
\end{abstract}



\section{Introduction}\label{sec:intro}

Tracing the evolution of galaxies and their mass assembly across cosmic time back to the Epoch of Reionization \citep[EoR, $z>6$;][]{Robertson2010,Barkana2001} is the current frontier of extragalactic astrophysics.
Several open questions are still waiting to be answered, including:
{\em How were galaxies initially formed?}
{\em How were early galaxies enriched with dust and metals?}
{\em What governs the star formation process in galaxies?}
{\em What is the origin of the star-forming main-sequence?}
{\em How do actively star-forming systems end up as quiescent galaxy populations seen today?}
We know that, overall, today's massive galaxies become rapidly chemically enriched and build their structure and stellar mass during the EoR and the early growth phase \citep[$z>4$;][]{maiolino19,curti23,faisst25b}, while cosmic noon ($z=1-3$) highlights the peak of the cosmic star formation rate (SFR) density and supermassive black hole (SMBH) growth \citep[related to active galactic nuclei, AGN;][]{madau14}. Subsequently, galaxies become quiescent systems \citep{peng10,ilbert10,Williams2021,Hartley2023} and the first quiescent-dominated galaxy clusters form at $z\approx2$ \citep{willis2020,finner2025}, the peak of cosmic noon.
Refining this picture through a complete multi-wavelength study of the different components (stars, dust, and gas) of galaxies on different spatial scales is the goal of current galaxy formation studies.

%%% FIGURE: 2 PANEL WITH FULL MOSAIC AND RMS MAP  %%%%
\begin{figure*}[t!]
\centering
\includegraphics[angle=0,width=1\textwidth]{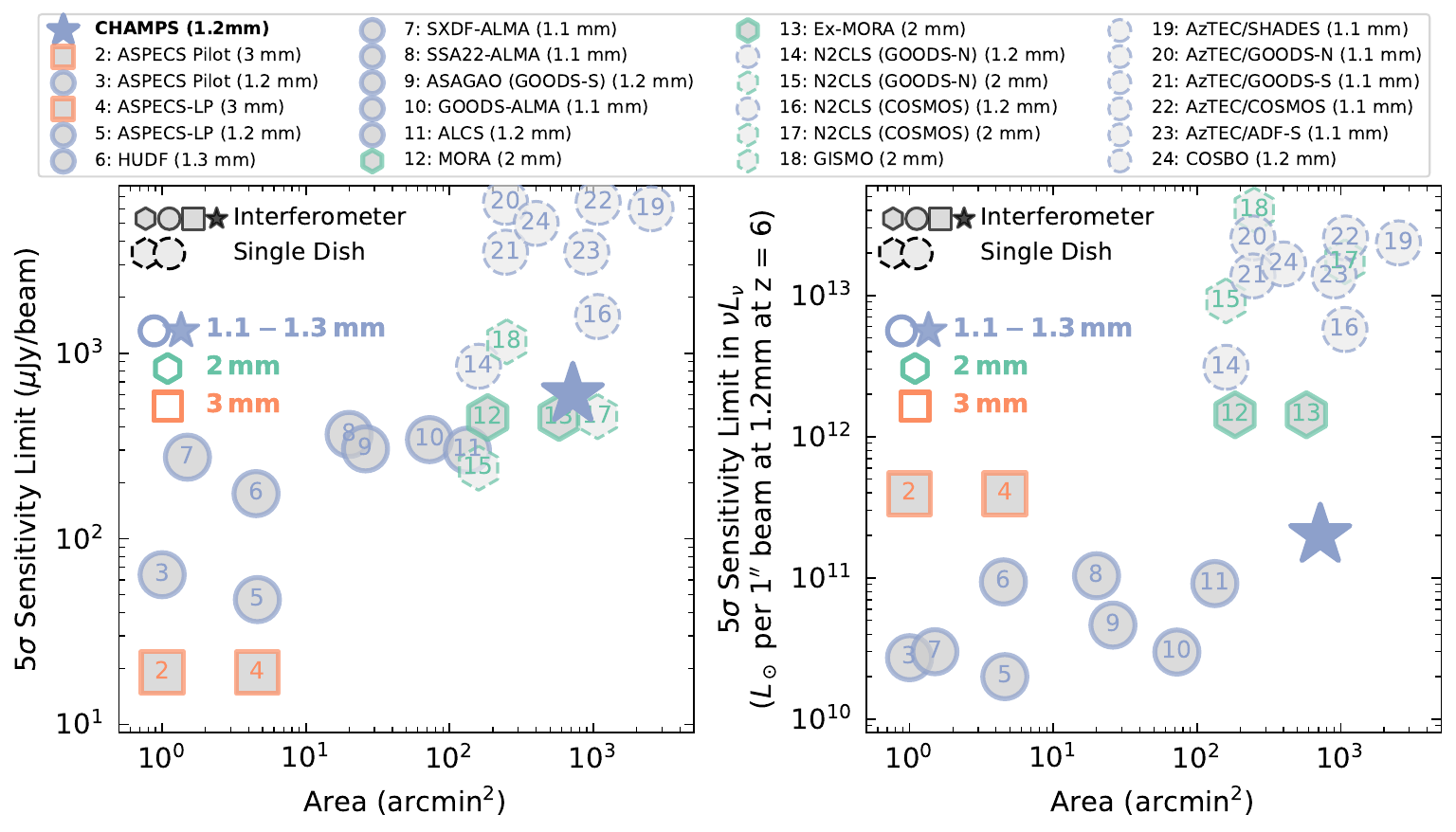}\vspace{-2mm}
\caption{
{\em Left:} Summary of sensitivity ($5\sigma$ per beam) and area coverage of various interferometric (solid) and single dish (dashed) sub-mm blind surveys in different fields (see numbers legend) at $1.1-1.3\,{\rm mm}$ (blue circles), $2\,{\rm mm}$ (green hexagons), and $3\,{\rm mm}$ (orange squares). \champs~is indicated as a star. The surveys include: ASPECS (Pilot and LP) \citep{Walter2016,Aravena2016,Aravena2020},
HUDF-ALMA \citep{Dunlop2017},
SXDF-ALMA \citep{Tadaki2015},
SSA22-ALMA \citep{Umehata2018,Umehata2017},
ASAGAO \citep{Hatsukade2016,Hatsukade2018},
GOODS-ALMA \citep{Franco2018,GomezGuijarro2022},
ALMA Lensing Cluster Survey \citep[ALCS;][]{Fujimoto2024},
GISMO \citep{Staguhn2014,Staguhn2014a,Magnelli2019},
AzTEC/SHADES \citep{Austermann2010,Michalowski2012},
AzTEC/GOODS-N \citep{Perera2008},
AzTEC/GOODS-S \citep{Scott2010},
AzTEC/COSMOS \citep{Scott2008,Aretxaga2011},
AzTEC/ADF-S \citep{Hatsukade2011},
(Ex-)MORA \citep{Casey2021,Long2026},
COSBO \citep{Bertoldi2007}, and
NIKA2 Cosmological Legacy Survey \citep[N2CLS;][]{Bing2023,Bethermin2026}.
{\em Right:} Same as left panel but for a $5\sigma$ luminosity ($\nu L_\nu$) limit normalized to a $1\arcsec$ beam and $1.2\,{\rm mm}$ at $z=6$.
\label{fig:overview}
}%\vspace{-3mm}
\end{figure*}
%%%%%%%%%%%%%%%%%%%%%%%%%%%%%%%

Out of these components, particular attention is being paid to the life cycle of dust, which plays an important role in galaxy evolution. Dust is created by a variety of mechanisms linked to star formation, supernovae, and interstellar medium (ISM) gas-phase metallicity \citep[e.g.,][]{Draine2003,asano2013dust-260,Schneider2024,Galliano2018}.
The detailed study of dust is not only of interest to understand its formation and properties \citep[such as grain sizes and charge, e.g.,][]{Mathis1977,Draine2001,Parente2026}, but also to interpret many basic astrophysical observables, such as to robustly measure the stellar mass \citep{Stringer2009,salim20} or to constrain the contribution of dust-obscured star formation to the total cosmic SFR density in the early universe \citep[][]{Casey2014,Traina2026}. 

It became clear that the dust-obscured universe needs to be studied; the presence of dust also introduces biases in the selection completeness of galaxy samples by obscuring some of the systems and reducing their flux below the detection threshold.
\citet{Bouwens2009} \citep[see also][]{Smail1997,Hughes1998} show that only $20\%$ of the light of $z=2$ galaxies is emitted at rest-frame UV and optical wavelengths, while the remaining $80\%$ can only be detected in the mid- and far-IR \citep{Bouwens2020}. At $z=5$, still half the light is expected to be missed in UV-based studies \citep{Bouwens2009,fudamoto20,inami22,Martis2025}.
In line with that, other studies find that about $40\%$ of the SFR density at $z>4$ is missed by UV-selected samples \citep[e.g.,][]{Zavala2021,Casey2014}.
This substantial gap in knowledge is further confirmed by recent investigations with the Atacama Large (Sub)Millimeter Array (ALMA).

%%% FIGURE: 2 PANEL WITH FULL MOSAIC AND RMS MAP  %%%%
\begin{figure*}[t!]
\centering
\includegraphics[angle=0,width=1\textwidth]{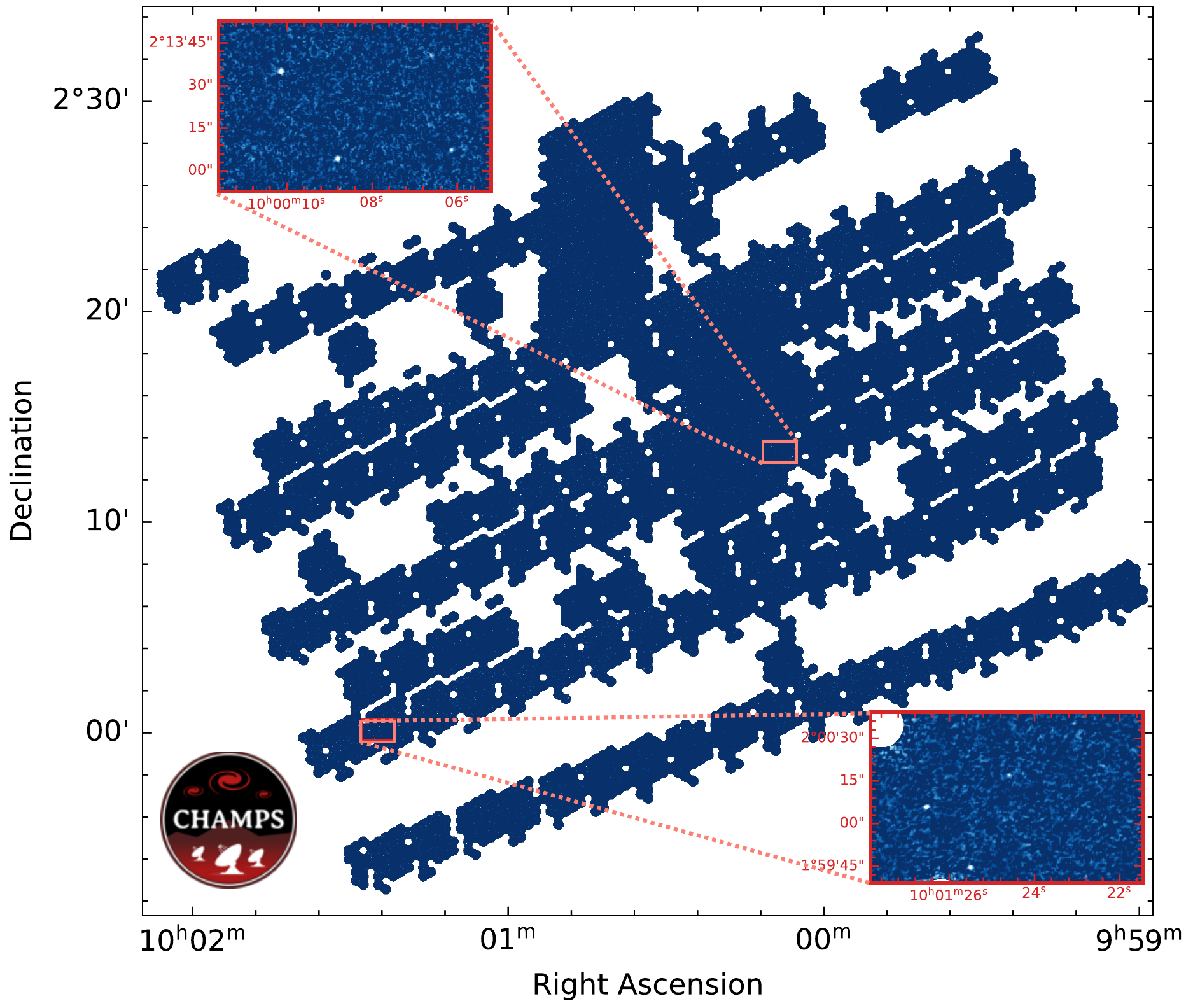}\vspace{-2mm}
\caption{
Complete \champs~SNR mosaic with two random zoom-ins at ($\rm 10^h00^m08^s$, $\rm +2^d13^m22^s$) and ($\rm 10^h01^m25^s$, $\rm +2^d00^m09^s$), respectively, highlighting a few of the sources detected. Each zoom-in spans a size of $1^{\prime} \times 1.6^{\prime}$. Note the odd shape of \champs~coverage, which follows the pointings of JWST/MIRI imaging from COSMOS-Web and PRIMER-COSMOS (see also Figure~\ref{fig:coverage}, right).
\label{fig:map}
}%\vspace{-3mm}
\end{figure*}
%%%%%%%%%%%%%%%%%%%%%%%%%%%%%%%

Due to its sensitivity and spatial resolution, ALMA can explore the far-IR continuum and line emission of high-redshift galaxies close to or during the EoR. This has led to large ALMA programs such as the {\it ALMA Large Program to Investigate \Cii~at Early Times} \citep[ALPINE;][]{lefevre20,bethermin20,faisst20} or the {\it ALMA Reionization Era Bright Emission Line Survey} \citep[REBELS;][]{bouwens22}, with the goal to target galaxy populations at specific redshifts in the early universe.
These surveys observed the largest samples of typical main-sequence star-forming galaxies at $z\approx4-8$ and aim to constrain their dust and gas properties via the measurement of far-IR emission lines (such as \Cii$_{158\,{\rm \mu m}}$ and \oiii$_{88\,{\rm \mu m}}$) and the far-IR dust continuum.
The results from these surveys have shown that such galaxies, although UV-selected, may already be significantly dust enriched \citep{faisst22,fudamoto20,pozzi21,Gruppioni2020,inami22,Mauerhofer2023,burgarella25,Inami2024,Algera2026}.

Even more curious is the serendipitous detection of galaxies in the ALMA beams of these {\em targeted} surveys \citep{Franco2018,fudamoto21,Gruppioni2020,Loiacono2021,Romano2020,vanLeeuwen2024}. Although these serendipitous sources are bright in \Cii$_{158\,{\rm \mu m}}$~emission or far-IR continuum, they remain undetected in Hubble space telescope optical imaging (and have thus been labeled ``near-IR dark'' galaxies), suggesting that a substantial population of dusty, high-redshift galaxies contributing to the dust-obscured cosmic SFR density may have been overlooked \citep[see also][]{Hughes1998}.
In addition to ALMA, other studies have focused on characterizing and finding such sources using deep radio data \citep{Talia2021,Gentile2024,Gentile2025,vandervlugt2023} or with the James Webb space telescope \citep[JWST;][]{Barrufet2023,PerezGonzalez2023,Williams2024,Kokorev2023,Gottumukkala2024}.
Thanks to the extended wavelength coverage and unprecedented sensitivity of JWST, most of these sources could be recovered, but at very faint rest-frame UV or optical fluxes.

The detailed optical and far-IR properties of typical galaxies derived from the ALPINE and REBELS samples and the measurements of the more detailed dust grain properties at lower redshifts \citep[most recently the spatially resolved polycyclic aromatic hydrocarbon (PAH) studies at $z\approx 1$; ][]{Wang2026,Lofaro2026,Donnan2026} are now informing new theoretical models of dust grain formation and evolution through cosmic time \citep[e.g.,][]{pozzi21,Narayanan2026,Osman25,Nanni2025,Donevski2026}.
A more complete sample of early dust obscured galaxies at high redshifts and robust measurements of their properties (such as their dust and stellar masses) and number densities would offer further strong constraints on the dust build-up in the early universe.

However, the total blind-survey-area enabled by targeted ALMA large programs is still too small (ALPINE and REBELS cover $\sim9\,{\rm arcmin^2}$ in total) to draw definite conclusions on the contribution of dusty sources to the global cosmic SFR density (hence mass growth) and on the number densities of the most dusty sources in the early universe.
Additionally, as shown by \citet{Loiacono2021} and \citet{fudamoto21}, clustering of galaxies may bias number counts (by increasing them) as only a small spatial area (typically $\sim10-20\arcsec$ corresponding to $\sim50-100\,{\rm kpc}$ at $z=5$) around the main targets are observed.

%%% FIGURE: SED AND SPATIAL COVERAGE  %%%%
\begin{figure*}[t!]
\centering
\includegraphics[angle=0,width=1.2\columnwidth]{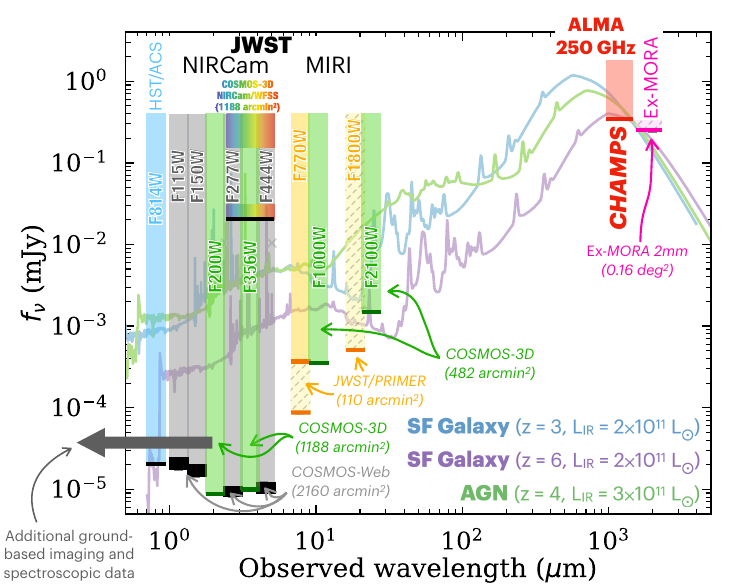}
\includegraphics[angle=0,width=0.9\columnwidth]{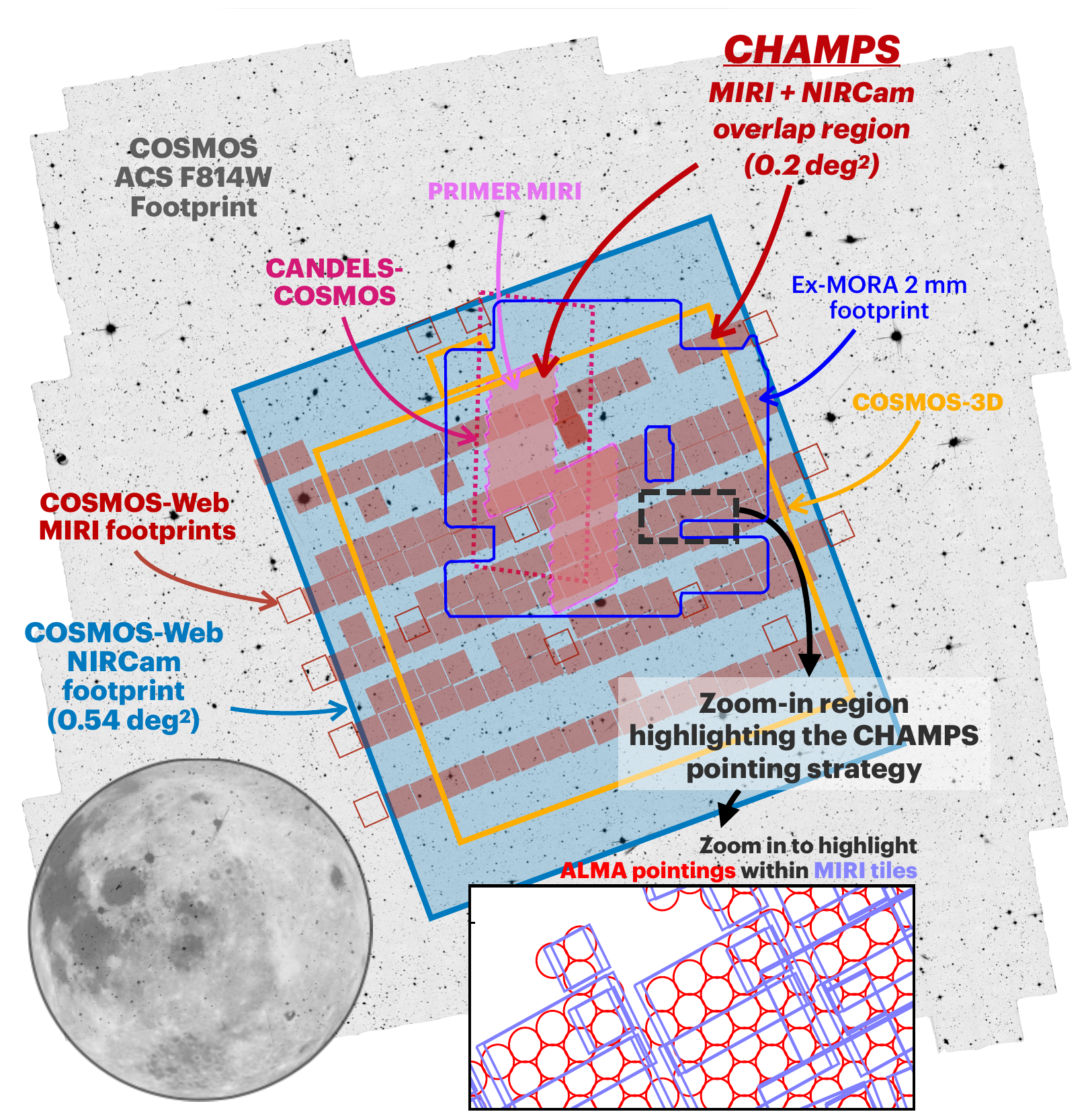}
%\vspace{6cm}
\caption{
{\em Left:} Overview of the spectral coverage showing \champs~(red, $1.2\,{\rm mm}$) as well as other observations from Hubble, JWST, and ALMA. We show example SEDs of two star-forming galaxies (at $z=3$ and $z=6$) and an AGN at $z=4$ with varying total IR luminosities as indicated in the legend.
{\em Right:} Overview of the \champs~sky coverage (red) compared to other observations from Hubble, JWST/NIRCam and MIRI, as well as ALMA. The inset focuses on a small region showcasing the pointing strategy of \champs~to cover all JWST/MIRI observations from COSMOS-Web, PRIMER-COSMOS, and COSMOS-3D. The background shows the COSMOS ACS/F814W footprint \citep{koekemoer07} with our Moon to scale.
\label{fig:coverage}
}%\vspace{-3mm}
\end{figure*}
%%%%%%%%%%%%%%%%%%%%%%%%%%%%%%%

In summary, these findings show that UV or optically-selected samples are not representative of the full population of dusty galaxies at high redshifts. 
A better way to pursue an unbiased study of the dust-rich galaxy population is by performing a blind survey at sub-mm wavelengths without target selection. 
Dedicated deep blind-field ALMA surveys were conducted in several deep fields (see Figure~\ref{fig:overview} and references in the caption), covering an area of $260\,{\rm arcmin^2}$ at $1.2\,{\rm mm}$ and $4.6\,{\rm arcmin^2}$ at $3\,{\rm mm}$, respectively.
Furthermore, recently single-dish blind-field observations with NIKA2 on the IRAM $30$-meter telescope were carried out at $1.2\,{\rm mm}$ and $2\,{\rm mm}$ over $1200\,{\rm arcmin^2}$ \citep{Bing2023,Bethermin2026}.
While these blind-field observations probe a complementary population of galaxies compared to targeted observations mentioned above, they still lack a sufficiently large area, or, in the case of large beam sizes of single-dish observations, render source identification difficult.

The only truly unbiased and comprehensive way to study the most dusty sources at early times, hence but the most stringent constraints on the early dust build-up, is by pursuing a large-area, untargeted blind-field survey with ALMA supported by deep JWST/MIRI observations.
This motivates the {\em COSMOS High-Redshift ALMA-MIRI Population Survey} (\champs; \#2023.1.00180.L, PI: Faisst), a new ALMA large program.
\champs~is currently the largest $1.2\,{\rm mm}$ blind-field survey to-date, covering a total of $0.2\,{\rm deg^2}$ (Figure~\ref{fig:map}).
In contrast to the recent {\em Mapping Obscuration to Reionization} \citep[MORA;][]{Casey2021,Zavala2021} and Extended (Ex)-MORA \citep{Long2026} ALMA blind-field surveys (covering an equal amount of area at $2\,{\rm mm}$ on the COSMOS field), \champs~was designed to cover co-spatially various JWST/NIRCam+MIRI observation (Section~\ref{sec:specs}) with complementary $1.2\,{\rm mm}$ observations over its $720\,{\rm arcmin^2}$.
This enables a joint analysis of JWST/NIRCam, NIRSpec, MIRI, and ALMA $1.2\,{\rm mm}$ data for a full characterization of the UV-to-IR spectral energy distributions (SEDs) of dust-obscured galaxies out to $z\approx7$, as well as to provide crucial observations to inform theoretical models of dust production at very high redshifts.

In this paper, we present the setup, science goals, and first science highlights of \champs.
In Section~\ref{sec:setup}, we provide a detailed description of the observational setup and the specific characteristics of the survey.
In Section~\ref{sec:science}, we delineate the scientific objectives of \champs~and showcase the observations. We conclude with Section~\ref{sec:summary}.
Throughout this work, we assume a $\Lambda$CDM cosmology (with $H_0 = 70\,{\rm km\,s^{-1}\,Mpc^{-1}}$, $\Omega_\Lambda = 0.7$, and $\Omega_{\rm m} = 0.3$). Magnitudes are given in the AB system \citep{oke74}. We use a Chabrier initial mass function \citep[IMF;][]{chabrier03} calibration to estimate stellar masses and SFRs.

\section{CHAMPS~Setup and Observations}\label{sec:setup}

\subsection{CHAMPS~Main Specifications}\label{sec:specs}

\champs's primary goal is to detect high-redshift dust-obscured galaxies, to 
{\em (i)} characterize their full UV to far-IR to radio SEDs to obtain a robust census of their properties over cosmic time, 
{\em (ii)} quantify their contribution to the total cosmic SFR density, 
{\em (iii)} inform models of early dust formation, and
{\em (iv)} perform a stacking analysis of fainter, optical to mid-IR detected galaxies to study average source properties across cosmic time.
With that in mind, \champs~obtained uniform ALMA $1.2\,{\rm mm}$ ($250\,{\rm GHz}$, band 6) observations at $0.5-1.2\arcsec$ resolution (configurations C-1 to C-4) over the largest possible area with deep JWST/NIRCam and MIRI (F770W) coverage. \champs~charged a total time of $143.5$ hours and includes $4800$ single pointings split in $32$ scheduling blocks (SBs, each containing 150 pointings).
Figure~\ref{fig:map} shows the full \champs~field with insets displaying a zoom-in to particular regions outlined by the red boxes.
In the following, we motivate the specifications of \champs.

\paragraph{Frequency Selection.~}~
Band 6 observations are optimal for the selection of dust-obscured $z>3$ galaxies \citep[e.g.,][]{Casey2014,Hodge2020}. Thanks to the negative $k-$correction and the typical dust SED shape, galaxies at $z=3-6$ are brightest at this frequency \citep[e.g.,][]{Blain2002}. This approach is similar to the MORA and Ex-MORA ALMA large programs, though due to their lower target frequency ($2\,{\rm mm}$, $147\,{\rm GHz}$, band 4) these programs are sensitive to even higher redshifts.
With band 6 observations, \champs~is $5-10\times$ more sensitive to galaxies at $z\sim5$ compared to the MORA survey. \champs~therefore optimally complements these other studies.
In addition, the combination of \champs~and MORA ($\sim50\%$ overlap) provides the unique opportunity to constrain the dust continuum slope to study dust emissivity and temperature. 
The left panel of Figure~\ref{fig:coverage} shows the SEDs of typical galaxies and an AGN at these redshifts together with the wavelength coverage from different other programs.

%%% FIGURE: RMS Map and BEAM SIZES  %%%%
\begin{figure}[t!]
\centering
\includegraphics[angle=0,width=1.05\columnwidth]{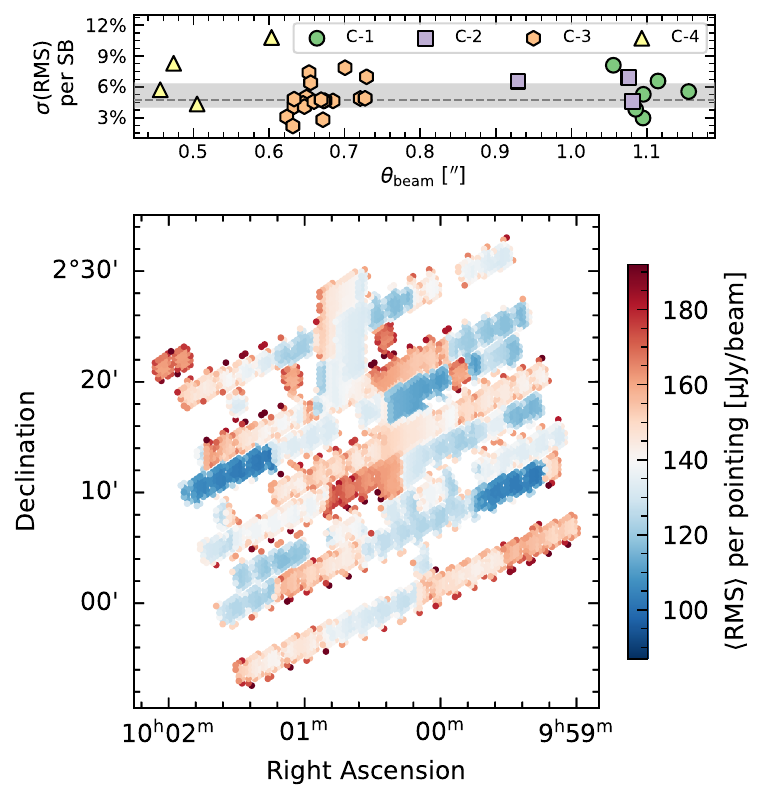}
\caption{
{\em Top:} Variations in the RMS per SB as a function of median beam size (per SB). The gray band shows the $\pm1\sigma$ spread. We find RMS variations of $3-6\%$ in regions of similar beam size. The symbols are color-coded by the respective ALMA configurations.
{\em Bottom:} Map of the average RMS per pointing. The large variations occur as observations are taken in different configurations, hence yielding different beam sizes.
Note that the final mosaic is constructed using a uniform synthesized beam of $1.1\arcsec\times1.1\arcsec$.
See \citet{Martinez2026} for more details on the RMS properties.
\label{fig:rmsmap}
}%\vspace{-3mm}
\end{figure}
%%%%%%%%%%%%%%%%%%%%%%%%%%%%%%%

\paragraph{Location and Ancillary Data.~}~
The \champs~area is chosen to leverage JWST/NIRCam and MIRI observations to find the most dust-obscured sources and at the same time characterize their properties from complete SED coverage.
This is motivated by a number of red sources detected by JWST \citep[e.g.,][]{Jermann2026,Akins2023}, which are potentially dust-obscured high-redshift galaxy systems.
\champs~is co-spatial with the NIRCam and MIRI coverage of the COSMOS-Web \citep[PID: $\#$1727, PIs: J. Kartaltepe \& C. Casey;][]{casey22},
PRIMER-COSMOS \citep[PID: $\#$1837, PI: J. Dunlop;][]{Dunlop2021,Donnan2024},
and COSMOS-3D \citep[PID: $\#$5893, PI: K. Kakiichi;][]{Kakiichi2024} JWST programs situated on the COSMOS field \citep{scoville07}.
COSMOS-Web provides coverage in F115W, F150W, F277W, and F444W \citep[NIRCam;][]{franco25}, and F770W \citep[MIRI;][]{harish25}. The PRIMER area provides additional F090W, F200W, F356W, and F410M (NIRCam) as well as F1800W (MIRI) imaging.
A significant part of \champs~is also covered by the COSMOS-3D JWST program \citep[PID: $\#$5893][]{Kakiichi2024}, adding deep F200W and F356W NIRCam coverage, F1000W and F2100W MIRI coverage, and NIRCam/WFSS slitless grism spectroscopy with the F444W cross filter.
The full field is also covered with HST/ACS F814W imaging \citep{koekemoer07} and partially with other HST filters covering the UV to optical wavelengths (e.g., from the CLUTCH program, $\#$17802, PI: Kartaltepe).
Thanks to being an equatorial field, COSMOS has been observed by many other space and ground-based telescopes covering magnitudes in frequency. Most of these are summarized in \citet{weaver22} and \citet{shuntov25}.
This includes X-ray observations from XMM-Newton \citep{Hasinger2007} and Chandra \citep{Civano2016,Marchesi2016} and radio observations from the MeerKAT MIGHTEE survey \citep[L-band at $1.3\,{\rm GHz}$ and $2.6\,{\rm GHz}$;][]{Jarvis2016,Hale2025}, VLA \citep[$1.4\,{\rm GHz}$, $3\,{\rm GHz}$, and $10\,{\rm GHz}$;][]{Schinnerer2007,schinnerer10,Smolcic2017,VanderVlugt2021}, LOFAR \citep[$144\,{\rm MHz}$;][]{Vardoulaki2026}, as well as far-IR measurements from SCUBA-2 $450\,{\rm \mu m}$ and $850\,{\rm \mu m}$ from STUDIES \citep{Wang2017} and S2COSMOS \citep{Simpson2019}. Archival ALMA data can be found in the A3COSMOS catalog \citep{Liu2019,Adscheid2024}, and additional sub-mm single-dish data is obtained by the IRAM 30-meter NIKA2 Cosmological Legacy Survey \citep[e.g.,][]{Bethermin2026,Bing2023}.
The right panel of Figure~\ref{fig:coverage} summarizes the different spatial coverage of \champs~and other surveys on the COSMOS field.

\paragraph{Uniform Coverage.~}~
To provide a robust statistical analysis of the most dusty galaxies in the early universe, \champs~requires a uniform selection function. The $4800$ \champs~pointings are therefore laid out at $2\times$ Nyquist sampling, i.e. a pointing center separation of $0.51093\times2\times{\rm FWHM}\approx24.66\arcsec$ of the primary beam.
This results in a uniform root-mean-square (RMS) noise characterization for observations in the same configuration.
We note that Nyquist sampling would increase the number of ALMA pointings to more than 17,000, rendering such a survey impractical.
Since the beam size (i.e., configuration that was used) varies across the \champs~field, the resulting RMS at a fixed restoring beam ($1.1\arcsec$) varies spatially across the map.
Figure~\ref{fig:rmsmap} shows the measured variation in RMS for groups of SBs with the same beam size ($\sim3-6\%$; top) and the full RMS map average over pointing (bottom).
Note that the edges of the coverage (specifically the MIRI-Lyot regions) have $\sim10\%$ worse noise properties due to the lack of coverage from nearby ALMA pointings.

\paragraph{Spatial Resolution.~}~
Sub-mm surveys of similar or larger area have been carried out by single-dish observatories. One example is the NIKA2 Cosmological Legacy Survey \citep[N2CLS;][]{Bing2023,Bethermin2026}, covering the GOODS-N and COSMOS fields with observations at $1.2$ and $2\,{\rm mm}$ with the IRAM 30-meter single-dish telescope. While the sensitivity and area is similar to \champs~($315\,{\rm \mu Jy/beam}$ and $91\,{\rm \mu Jy/beam}$ at $1.2\,{\rm mm}$ and $2\,{\rm mm}$, respectively, Figure~\ref{fig:overview}), the beam sizes of $11.6\arcsec$ and $18\arcsec$ at $1.2$ and $2\,{\rm mm}$, respectively, make source identification challenging \citep{CarvajalBohorquez2026}. With an average (approximately circular) beam size of $0.8\arcsec$ (and synthesized to a uniform $1.1\arcsec$), \champs~probes a different parameter space, allowing the clean identification of dusty high-redshift sources using ancillary JWST data. 
\champs~is taken in configurations C-1 to C-4, with most observations taken in configuration C-3 at a beam size of $\sim 0.7\arcsec$ (Figure~\ref{fig:rmsmap} top panel). However, there are some pointings ($\sim10\%$ of area) that were observed in C-1 at a significantly larger beam sizes of $\sim1.2\arcsec$.

%%% FIGURE: LIR(z) AND EMISSION LINES  %%%%
\begin{figure}[t!]
\centering
\includegraphics[angle=0,width=\columnwidth]{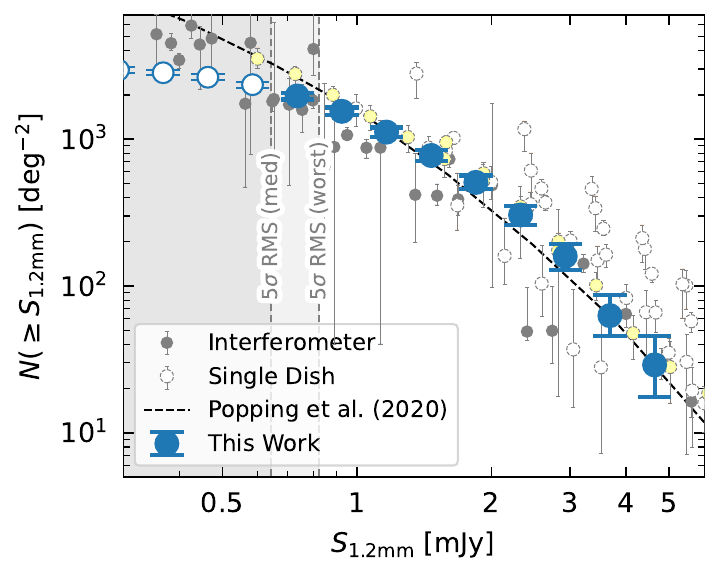}\vspace{-3mm}
\caption{
Total cumulative $1.2\,{\rm mm}$ number density counts obtained by \champs~(large blue circles). The curvature at fainter fluxes is due to sample incompleteness \citep[see][]{Martinez2026}. The $5\sigma$ limit (median and worst) is indicated. Error bars include Poisson and photometric uncertainties.
We show single dish (dashed circles) and
interferometric (solid gray circles) measurements (see references in text).
The yellow dashed circles highlight the single dish measurements from the N2CLS \citep{Bing2023,Bethermin2026,CarvajalBohorquez2026}.
The dashed line shows the semi-empirical model by \citet{Popping2020} including all galaxies at $z<7$.
\label{fig:sourcedensity}
 }%\vspace{-3mm}
\end{figure}
%%%%%%%%%%%%%%%%%%%%%%%%%%%%%%%

\subsection{\champs~Observations and Data Reduction}

Here, we give a brief overview of the observations and data reduction of \champs. For more details we refer to \citet{Martinez2026}.
The 4800 \champs~pointings have been observed in 32 different SBs, with integration times ranging from $54.4\,{\rm s}$ to $90.7\,{\rm s}$ per pointing. The resulting 5\%-95\% percentile RMS ranges from $97$ to $263\,{\rm \mu Jy/beam}$ with a median at $135\,{\rm \mu Jy/beam}$.
The observations were carried out between January and November 2024 in four $1.875\,{\rm GHz}$-wide spectral windows centered at $241$, $242.8$, $257$, and $258.8\,{\rm GHz}$, respectively. As noted above, configurations C-1, C-2, C-3, and C-4 were used resulting in beam sizes that vary from $0.46\arcsec$ to $1.16\arcsec$.

%%% FIGURE: LIR(z) SENSITIVTY AND COLOR  %%%%
\begin{figure*}[t!]
\centering
\includegraphics[angle=0,width=1\columnwidth]{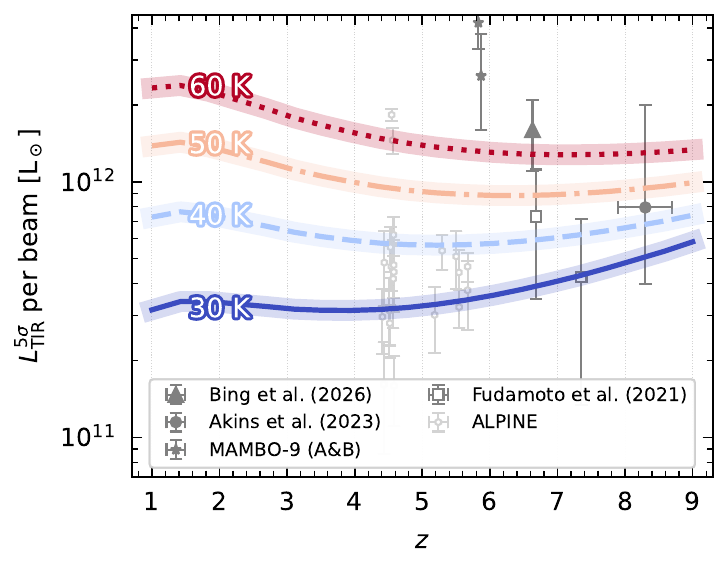}
\includegraphics[angle=0,width=1\columnwidth]{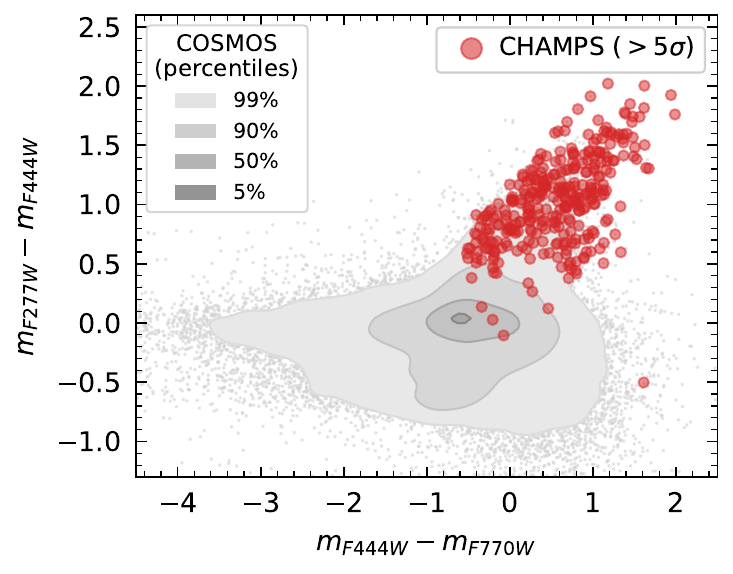}\vspace{-2mm}
\caption{
{\em Left:} Estimated $5\sigma$ total IR luminosity limits (per synthesized beam) as a function of redshift and dust temperature ($30\,{\rm K}-60\,{\rm K}$).
Also shown are measurements from ALPINE \citep[open circles;][]{bethermin20}, two serendipitously detected near-IR dark galaxies from REBELS \citep[empty squares;][]{fudamoto21}, one of the dust-obscured JWST/MIRI-detected massive galaxies from \citet{Akins2023} (solid circle), and a ALMA-detected near-IR dark galaxy from \citet{Bing2026} (filled triangle).
{\em Right:} Color-color parameter space probed by \champs~detections ($>5\sigma$, red) compared to the galaxies detected in COSMOS-Web at similar redshifts (gray, contours cover 5, 50, 90, and $99\%$ of galaxies). The majority of \champs~detections exhibit significantly redder colors than $\approx99\%$ of the galaxies detected with JWST. This figure is equivalent to the one presented in \citet{Zavala2026}.
\label{fig:sensitivity}
}%\vspace{-3mm}
\end{figure*}
%%%%%%%%%%%%%%%%%%%%%%%%%%%%%%%

For data reduction, the \texttt{tclean} method (adopting {\em natural} weighting) was used from the standard CASA-ALMA pipeline \citep[version \texttt{6.6.1.17};][]{Hunter2023}.
It was run on designated clusters maintained by the National Radio Astronomy Observatory (NRAO). To support continuity in the final \champs~mosaic, the full mosaic was reduced at once, however, the process was optimized by using the CASA \texttt{split} command to lower the total amount of data per reduction step and to reduce the total computation time.
Since we focus here on continuum imaging (the reduction of the spectral cubes will be discussed and presented in a forthcoming paper), frequency binning was applied (8 channel bins of $125\,{\rm MHz}$), which results in decreased noise properties.
Furthermore, the different beam sizes were taken care of by setting a uniform restoring beam size across the mosaic of $1.1\arcsec$ (equivalent to the worst beam size).
Finally, the different pieces were combined to a mosaic with final pixel resolution of $0.15\arcsec$ using the CASA \texttt{concat} command.
The primary beam correction was applied to the full mosaic using the CASA command \texttt{impbcor}.

The source detection was performed on the final signal-to-noise (SNR) mosaic based on their peak SNR. For blind source detection, purity rates of $96.9\%$ at $\rm SNR>5$ and $36.1\%$ at $4.5 < {\rm SNR} < 5$, respectively, were reached (computed from the number of positive and negative flux peaks).
Once cross-matched (within $r=1\arcsec$) to the COSMOS2025 catalog \citep{shuntov25}, purity rates of $98.7\%$ at $\rm SNR\ge5$ and $71.7\%$ at $4.5 < {\rm SNR} < 5$, respectively, were found.
The total survey completeness (evaluated with injection simulations) reaches $90\%$ at a $\sim1\,{\rm mJy}$ for unresolved sources.
Details of source detection, purity, and completeness calculations can be found in \citet{Martinez2026}.

Figure~\ref{fig:sourcedensity} shows the number density of $1.2\,{\rm mm}$-detected \champs~sources per square-degree (large blue circles). The errors include Poisson\footnote{The Poisson errors are computed without approximation using the Python function \textsc{poisson\_conf\_interval()} as part of the \texttt{astropy.stats} package.} and photometric uncertainties. These number counts are compared to the semi-empirical model by \citet{Popping2020} (model including all galaxies at $z<7$), as well as single dish \citep[dashed circles;][]{Scott2008,Perera2008,Scott2010,Austermann2010,Hatsukade2011,Lindner2011,Aretxaga2011} and
interferometric \citep[solid gray circles;][]{Chen2023,Fujimoto2016,Aravena2016,GomezGuijarro2022,GonzalezLopez2020,Hatsukade2016,Hatsukade2018,Scott2012,Umehata2017} measurements.
We find that most single dish measurements overestimate the number counts, which may be due to insufficient deblending, except for the measurements by the NIKA2 Cosmological Legacy Survey \citep[N2CLS, highlighted in yellow;][]{Bing2023,Bethermin2026,CarvajalBohorquez2026}.
We note that the number counts start to turn over around $1\,{\rm mJy}$, which is where the survey becomes less complete \citep[for details see][]{Martinez2026}.

\subsection{Program Deliverables}\label{sec:deliverables}
The \champs~collaboration is planning to provide several data products to the community after the acceptance of the respective papers. These data products include the fully reduced continuum and spectral line map mosaic as well as source detection catalogs, total continuum and line flux measurements, and completeness statistics. 
Furthermore, we plan to publish ancillary data for each \champs-detected source in the form of image cutouts as well as their full best-fit SEDs obtained by various fitting programs based on the \champs~as well as ancillary photometry from UV to radio wavelengths. The latter includes several derived quantities such as stellar masses, dust masses, molecular gas masses derived from the sub-mm continuum, and morphological classifications.

\section{CHAMPS Science}\label{sec:science}

Since the advent of JWST, our understanding of the earliest galaxies and their morphological and chemical characteristics, reaching up to and into the EoR, has improved substantially. Nonetheless, these investigations (particularly at the highest redshifts) still rely primarily on rest-frame UV and optical observations. Although dedicated ALMA programs like ALPINE and REBELS deliver complementary constraints on the far-IR SEDs for a subset of these sources at specific redshifts, their strategy of focusing on previously identified (UV-bright) sources introduces distinct selection biases. 

\champs~provides a crucial blind-sky survey at $1.2\,{\rm mm}$ to complement the observations by JWST and other UV, optical, IR, sub-mm, and radio facilities on the COSMOS field \citep{weaver22}.
\champs~opens the possibility to enhance our knowledge of the most dusty sources back to the EoR, understand the missing star formation towards cosmic reionization, and perform statistical analyses of gas and dust of JWST-detected sources. The additional sub-mm anchor point from \champs~is crucial for finding the most dusty sources, to derive robust redshifts and physical properties for them, and to constrain galaxy formation models.

The following sections outline the most important \champs~science goals in more detail and present some initial findings.

%%% FIGURE: BETTER DUST MASSES AND REDSHIFTS  %%%%
\begin{figure*}[t!]
\centering
\includegraphics[angle=0,width=0.97\columnwidth]{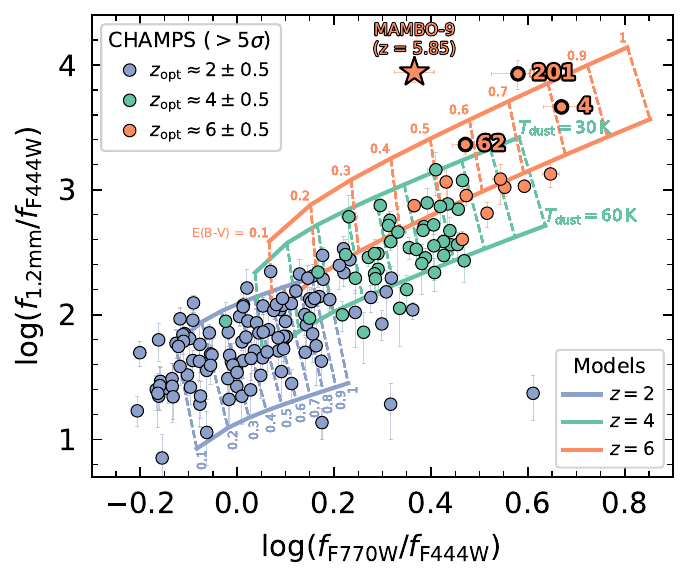}
\includegraphics[angle=0,width=1.03\columnwidth]{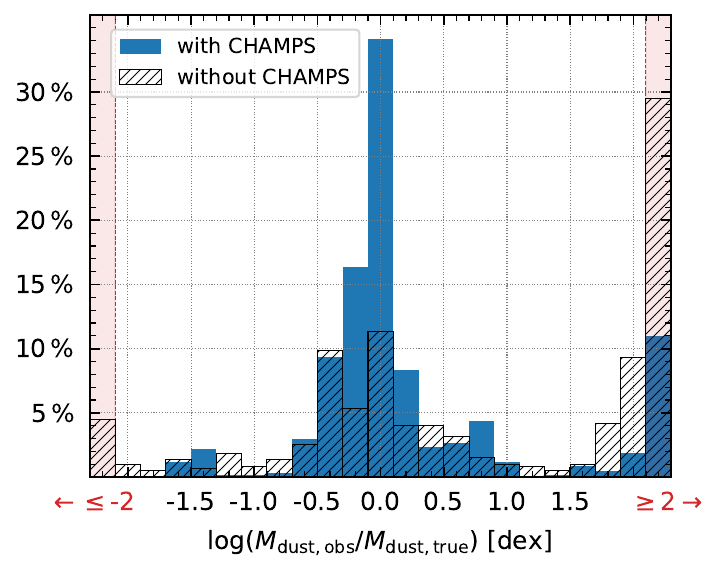}\vspace{-2mm}
\caption{
{\em Left:} JWST vs. ALMA $1.2\,{\rm mm}$ color-color space. Grids show models derived with \texttt{CIGALE} at different redshifts and assuming different E(B$-$V) values and dust temperatures ($30\,{\rm K}$ and $60\,{\rm K}$). Symbols show \champs~detections at $>5\sigma$ (cross matched with COSMOS2025) color-coded by similar optical photometric redshift ranges.
CHAMPS-4, CHAMPS-37 \citep[MAMBO-9 at $z=5.85$;][]{Bertoldi2007,Jin2019}, CHAMPS-62, and CHAMPS-201 are among the most dusty and highest redshift candidate galaxies in the sample.
{\em Right:} Demonstration of the importance of \champs~$1.2\,{\rm mm}$ photometry for the derivation of dust masses (see text for description of simulation and fitting). The simulated galaxies are fit with \texttt{CIGALE} using the full COSMOS2025 \citep{shuntov25} photometry with (blue) and without (black hatched) \champs~data. With \champs, the dust masses can be recovered within $0.5\,{\rm dex}$ for $>80\%$ of the simulated galaxies. If omitting \champs~photometry, this fraction drops to $<40\%$ (with $35\%$ off by more than two orders of magnitudes). 
\label{fig:dustmass}
}%\vspace{-3mm}
\end{figure*}
%%%%%%%%%%%%%%%%%%%%%%%%%%%%%%%

\subsection{Identifying the Most Dusty Sources in the Early Universe}\label{sec:sciencegoal1}

Understanding the nature of dusty sources revealed by ALMA and JWST is extremely important \citep[e.g.,][]{Schneider2024}. Recent discoveries of galaxies with large dust reservoirs at, or even before, the EoR may challenge the long-standing assumption that these objects host only minimal amounts of dust \citep{Mitsuhashi2026,Rodighiero2026}.
Specifically, near-IR dark galaxies are intriguing \citep{Franco2018,fudamoto21,Gentile2025,Gruppioni2020,Loiacono2021,Williams2024}. Most of them are found serendipitously as by-products of targeted large ALMA surveys such as ALPINE and REBELS, but some are specifically selected through radio observations \citep{Gentile2024}. Notable examples are two dust continuum-detected $z\approx7$ galaxies in the REBELS ALMA field-of-view without an optical counterpart \citep{fudamoto21} or two extremely red JWST/MIRI-detected massive ($\rm M_\star>10^{10}\,{\rm M_\odot}$) $z\approx8$ galaxies \citep{Akins2023}.
Similarly intriguing is the dearth of dusty galaxies at $z\gtrsim8.5$, suggesting substantial dust production through supernovae and dust growth in the ISM between $z\sim 7.5-8.5$ \citep[e.g.,][]{Bakx2026,Burgarella2026,Algera2025}.

%%% FIGURE: CUTOUTS  %%%%
\begin{figure*}[t!]
\centering
\includegraphics[angle=0,width=2.1\columnwidth]{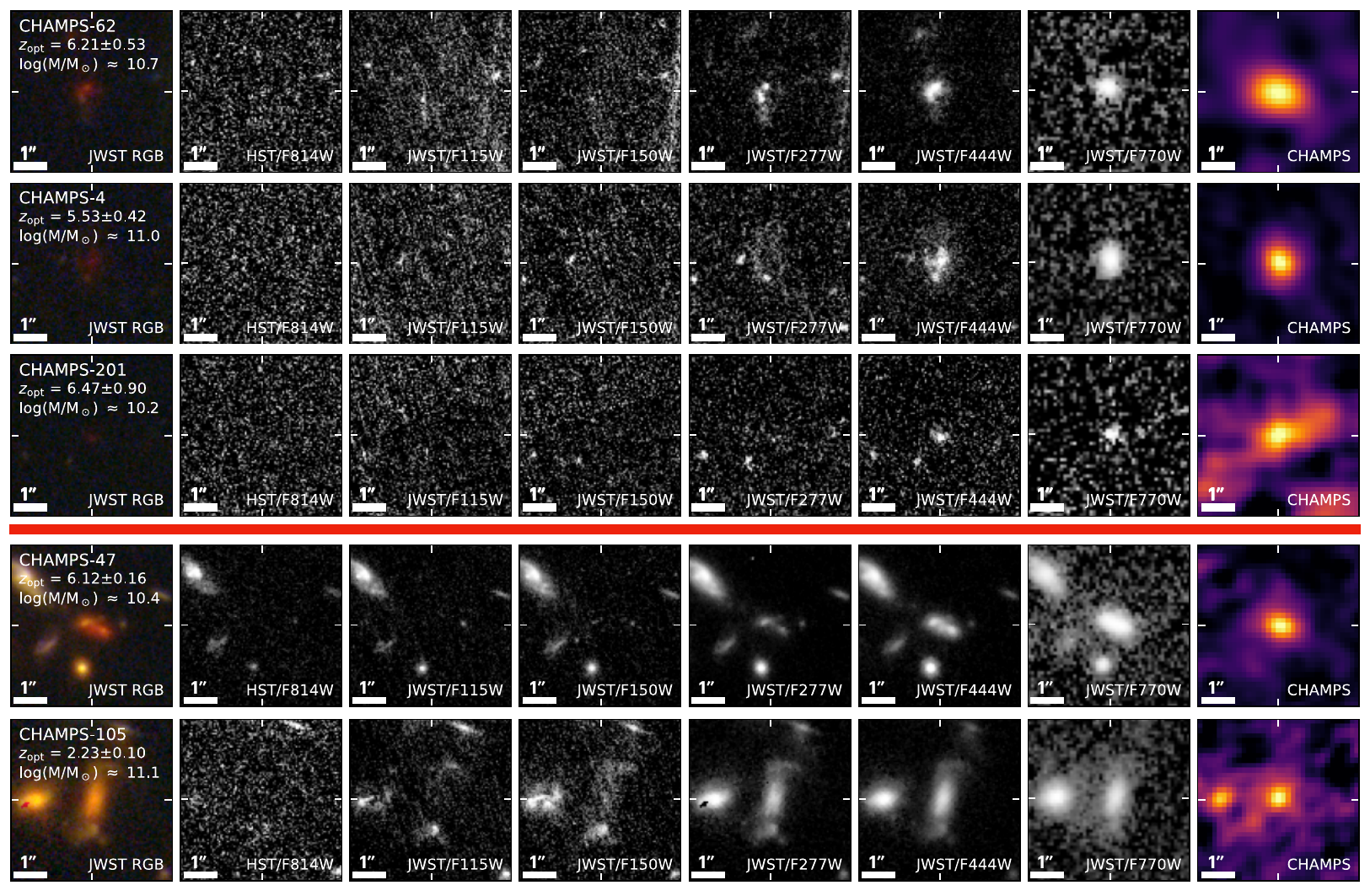}
\caption{
Cutouts of five example galaxies detected in \champs~(all at $>5\sigma$). The top three show the top three reddest (and highest redshift) sources (see also Figure~\ref{fig:dustmass}). The bottom two show lower redshift detections. Shown are the JWST/NIRCam RGB images (F150W-F277W-F444W; left), the JWST/NIRCam$+$MIRI cutouts (middle panels), and the \champs~$1.2\,{\rm mm}$ cutout (right). Optical redshifts (from COSMOS2025) are indicated on the left panels as well as their stellar masses. Optical$+$\champs~redshifts and stellar masses are indicated on the right panels. Note the heavy dust attenuation of these galaxies, making them essentially invisible in Hubble optical imaging.
\label{fig:cutouts}
}%\vspace{-3mm}
\end{figure*}
%%%%%%%%%%%%%%%%%%%%%%%%%%%%%%%

Targeted ALMA surveys only draw a pencil-beam view of what may be a much larger, unexplored population (ALPINE and REBELS combined cover less than $10\,{\rm arcmin^2}$).
\champs~finds such extremely dusty sources thanks to its large area coverage and its homogeneity, which allows the construction of a well-defined selection function to study their number densities. Its combination with NIRCam and MIRI imaging provides an unbiased view on the dusty early universe.
Specifically, \champs~aims to answer questions such as:
{\em How many of such extremely dust obscured sources do exist?}
{\em What is their contribution to the total cosmic SFR density at high redshifts?}
{\em What is the amount of dust produced and how do massive galaxies grow in the early universe?}

The left panel of Figure~\ref{fig:sensitivity} shows the final \champs~$5\sigma$ per-beam sensitivity limit of the total IR luminosity as a function of redshift.
Note that these derived IR luminosity limits depend strongly on the assumed dust temperature \citep[$S_\nu \propto T_{\rm dust}^{-3.5}$; e.g.,][]{Casey2009,Blain2002} as shown by the colored lines.
To derive the sensitivity curves, we used a modified black body model \citep{casey12,Blain2003} with fixed $\beta_{\rm dust} = 1.8$ (emissivity index), $\alpha=2$ (mid-IR slope), $\lambda_0=200\,{\rm \mu m}$ (wavelength of unity optical depth), and a range of dust temperatures including $30$, $40$, $50$, and $60\,{\rm K}$. These values are consistent with recent studies at high redshifts \citep[e.g.,][]{faisst20,sommovigo22a}.
The far-IR modified black body SEDs were normalized to the \champs~median $5\sigma$ RMS per beam and the total IR luminosity was derived subsequently by integration over $8-1000\,{\rm \mu m}$.
To put these limits into context, we show in the same figure measurements from the ALPINE sample \citep{bethermin20}, two near-IR dark galaxies found in REBELS \citep{fudamoto21}, an ALMA-detected near-IR dark galaxy \citep{Bing2026}, one of the two red massive JWST/MIRI-detected galaxy \citep{Akins2023}, and MAMBO-9, a pair of massive dusty star forming galaxies at $z=5.85$ \citep{Bertoldi2007,Aretxaga2011,Casey2013,Akins2026}.
\champs~is sensitive to galaxies similar to the ones by \citet{fudamoto21}, \citet{Akins2023}, and \citet{Bing2026}, which are generally more dust obscured than the galaxies in the ALPINE and REBELS samples.

The right panel of Figure~\ref{fig:sensitivity} presents the JWST color–color parameter space (spanned by F277W, F444W, and F770W) probed by \champs. The $>5\sigma$ \champs~detections \citep[see][]{Martinez2026} exhibit significantly redder colors than $99\%$ of the galaxies detected with JWST/NIRCam$+$MIRI in the COSMOS2025 catalog \citep{shuntov25}. This highlights the new region of parameter space accessed by \champs, which has not previously been investigated with sub-mm observations, as further discussed in \citep{Zavala2026}.

\champs~has already shown its potential by detecting a massive $\log{({\rm M/M_\odot})} \approx 11.3$ and $\rm L_{\rm TIR} \approx 6\times10^{12}\,{\rm L_\odot}$ galaxy spectroscopically confirmed to be at $z = 5$  \citep[ERD-1;][]{Gentile2024a}.
Deriving the robust number density estimates of such sources will be for the first time possible with \champs~thanks to its large area coverage. Depending on (the highly uncertain) models and based off previous small number statistics, we expect $10-30$ of them in the \champs~survey volume.

The combination of JWST and ALMA provide additional opportunities to directly find such dusty galaxies. One example of a new selection method involving NIRCam and MIRI photometry is demonstrated in \citet{Zavala2026}.
In addition, the left panel of Figure~\ref{fig:dustmass} shows a different selection method based on the color-color diagram involving the F444W$+$F770W JWST filters and the \champs~$1.2\,{\rm mm}$ data.
The grids show models for different redshifts created with \texttt{CIGALE} \citep{boquien19,burgarella25}. The models assume different dust attenuations and dust temperatures coupled to a modified black body far-IR SED. These models show how galaxies separate on this diagram; by redshift in the $y$-direction and by dust attenuation in the $x$-direction.
The symbols show \champs~$>5\sigma$-detections cross matched to the COSMOS2025 catalog and color-coded by their optical photometric redshift derived from JWST and ground-based data.
From this figure, we can select the top three of the most dusty high-redshift candidates (CHAMPS-4, CHAMPS-62, and CHAMPS-201). Note that the {\em star} is MAMBO-9 \citep[see][]{Akins2026}, CHAMPS-37, which is also included in the top four dust-obscured galaxies. MAMBO-9 is spectroscopically confirmed at $z=5.85$ \citep{Jin2019}.

The top three panels in Figure~\ref{fig:cutouts} show cutouts of these three interesting sources. Show are data from HST, JWST, and \champs. The photometric redshifts derived from optical to near-IR COSMOS2025 photometry ($z_{\rm opt}$) and \champs~data ($z_{\rm champs}$) suggest that two of these galaxies live in the EoR. At masses of $\rm \log(M/M_\odot) > 10.4$ and undetected in JWST/F150W (and two even in JWST/F277W), these galaxies are likely some of the most massive, most dust-obscured EoR galaxies ever detected.

The figure also shows two galaxies at lower redshifts. CHAMPS-47 originally at $z_{\rm opt}\approx6.12\pm0.16$ (from the COSMOS2025 catalog) was recovered at $z_{\rm champs}\approx3.00\pm0.25$. CHAMPS-105 originally at $z_{\rm opt}\approx2.23\pm0.10$ was recovered at $z_{\rm champs}\approx3.71\pm0.29$ including the \champs~data.
These discrepancies may be due to the inclusion of the $1.2\,{\rm mm}$ data, which is more discussed in a later publication. 

It is apparent from Figure~\ref{fig:cutouts} that the majority of \champs-detected sources are almost entirely undetected in the NIRCam/F150W observations and would consequently remain unseen in Hubble near-IR imaging (e.g., F160W) prior to JWST. They therefore represent prototypical examples of so‑called near‑IR dark galaxies.
This is to show the potential of \champs~in finding and characterizing the most dusty sources in the early Universe. The additional \champs~photometry point is crucial to measure their redshifts and other physical properties as shown in the next section.
A follow-up paper will focus on the physical properties of these intriguing sources as well as on the derivation of robust photometric redshifts including the \champs~data in comparison to optical photometric redshifts.

%%% FIGURE: MIRI, MEERKAT, CHAMPS, SEDs  %%%%
\begin{figure*}[t!]
\centering
\includegraphics[angle=0,width=2.13\columnwidth]{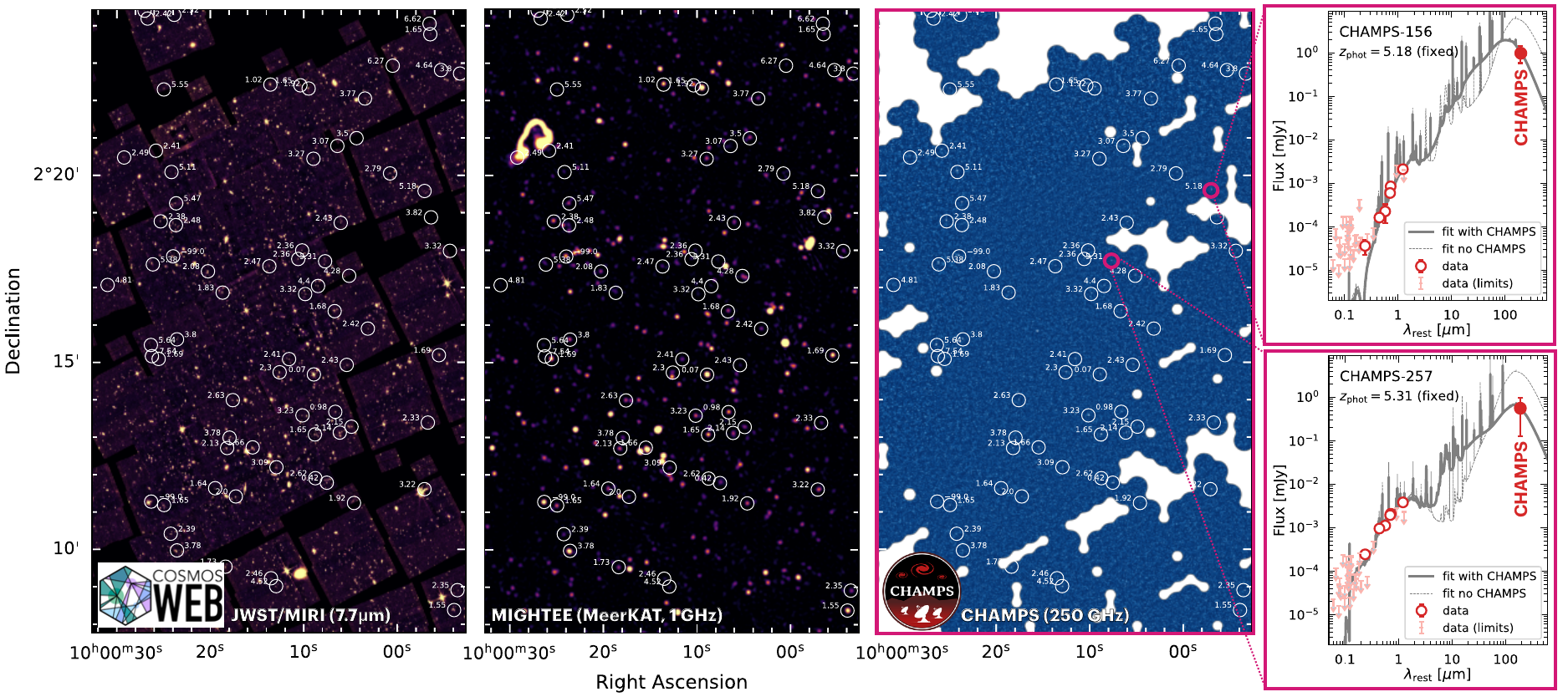}
\caption{
{\em Left three panels:} Large area cutouts ($10\arcmin\times17\arcmin$) centered on the JWST PRIMER-COSMOS area including JWST/MIRI $7.7\,{\rm \mu m}$ (COSMOS-Web and PRIMER-COSMOS; left), MeerKAT $1\,{\rm GHz}$ (MIGHTEE; center), and \champs~$1.2\,{\rm mm}$ (right). The circles indicate $>5\sigma$ detected \champs~sources and their photometric redshifts from the COSMOS2025 catalog.
{\em Right panels:} \texttt{CIGALE} fits including the COSMOS2025 and \champs~photometry to two example galaxies (CHAMPS-156 and CHAMPS-257) at photometric redshifts of $z=5.18$ and $z=5.31$, respectively. The fits including (solid) and omitting (dashed) \champs~data are shown.
\label{fig:mirimap}
}%\vspace{-3mm}
\end{figure*}
%%%%%%%%%%%%%%%%%%%%%%%%%%%%%%%

\subsection{Characterization of the Full Multi-Wavelength SEDs of Dusty Sources}\label{sec:sciencegoal2}

Dust obscured galaxy systems often have unconstrained SEDs. This is typically because either the rest-frame optical and visible regime is not detected, or the sources lack coverage in the far-IR.
Yet, having access to the complete SED is essential for deriving robust physical properties of these systems, including their redshifts, baryon masses (in stars, dust, and gas), dust-obscured SFRs, as well as dust temperatures and emissivities.
These properties, in turn, are key to understanding the intrinsic nature of dusty sources and the evolutionary stage they inhabit by measuring their star formation, stellar and gas mass, and depletion times.

In addition to the detection of dusty galaxies in the early universe (Section~\ref{sec:sciencegoal1}), the joint coverage of JWST/NIRCam$+$MIRI and \champs~sub-mm observations provides unique multi-wavelength SED measurements to better understand the physical properties of these sources. Furthermore, the high resolution JWST imaging up to $7.7\,{\rm \mu m}$ (and more redward in the PRIMER-COSMOS area thanks to MIRI/F1800W) allow for a robust association of sub-mm detections to near-IR sources and to examine their rest-frame optical morphology (Figure~\ref{fig:cutouts}).
With this, \champs~will help answer a series of questions, including:
{\em What are the physical and structural properties of dusty high-redshift sources?}
{\em What are the most effective ways to select the most dust-obscured systems?}
{\em What are the optical to far-IR properties of dust-obscured X-ray detected sources?}

To answer these questions, \champs~provides an important anchor point at $1.2\,{\rm mm}$ for SED fitting. This photometry point breaks degeneracies in the redshift estimation \citep[ruling out low-redshift solutions by providing constraints on the location of the far-IR peak, e.g.,][]{Battisti2019,Williams2024}, provides key measurements such as molecular gas masses \citep[by covering the Rayleigh-Jeans part of the IR spectrum, e.g.,][]{scoville13} and dust-obscured star formation rates, and improves many physical parameter estimates (including dust masses).

We demonstrate the additional constraining power of the \champs~photometry by using \texttt{CIGALE} to generate mock galaxy SEDs, then fitting them with and without the \champs~data.
We simulate $600$ models at $3 < z < 7$ with different radiation field parameters (using the updated far-IR parameterization of \citealt{Draine2014}) and $A_{\rm V}$ values \citep[following][]{calzetti00}\footnote{\texttt{CIGALE} enforces UV–IR energy balance, hence these parameter ranges imply different total dust masses.}. For each model, we predict the $1.2\,{\rm mm}$ flux and the fluxes in all COSMOS2025 photometric bands. We then re-fit the model photometry, with and without \champs, using a larger, comprehensive template library in \texttt{CIGALE}.
The right panel of Figure~\ref{fig:sensitivity} shows that with the inclusion of the \champs~photometry dust masses can be recovered within a factor of three ($0.5\,{\rm dex}$) of the true values for $>80\%$ of simulated galaxies at any sampled redshift. Omitting \champs~reduces this fraction to $<40\%$ and in addition $35\%$ are off by more than two orders of magnitude. This highlights the importance of even a single sub-mm data point for deriving dust masses, one of the key components of galaxies.

According to models by \citet{Casey2018}, we expect $\sim1200$ dusty sources to be detected at $3\sigma$ at $1.2\,{\rm mm}$ by \champs~over a redshift range out to the EoR. These number predictions are highly sensitive to the proportion of dust-obscured SFR and dust temperature. \champs~aims to constrain these models as one of its main objectives.
In addition, we expect $\sim500$ X-ray detected AGN \citep{Marchesi2016} to be covered by \champs~(most at $z>1$), approximately half of which are predicted to be bright enough to be detected individually \citep[e.g.,][]{Lanzuisi2017}.
By adding $1.2\,{\rm mm}$~observations to the available photometry, \champs~provides more robust measurements and more stringent constraints on stellar, dust, and molecular gas masses, as well as redshifts for all of these dusty sources over a redshift range out to $z=7$. Even non-detections provide valuable information and upper limits on some of these parameters.
The robust stellar masses derived in conjunction with JWST/MIRI rest-frame optical/near-IR photometry constrain the mass growth and characterize the relation between stellar mass and obscured fraction of SFR at high redshifts \citep{Whitaker2017,fudamoto20,inami22,Algera2023}.

The joint coverage with other sub-mm observations (including NIKA2 and Ex-MORA) provide multi-wavelength continuum constraints on dust temperatures and emissivities, hence also more robust measurements of the total IR luminosities \citep[which inherently depend on the assumption of dust temperature and emissivity, e.g.,][]{faisst17b}.
We note that, as shown by \citet{Silverman2026}, the far-IR luminosity as traced by \champs~still provides an unbiased view on the dust-obscured SFR even in bright quasars, as their radiation mostly affects the warmer mid-IR part of the SED.
Furthermore, radio data from MeerKAT (which covers the full \champs~footprint as part of the MIGHTEE program) allows to extend the SED coverage to the GHz regime. Several near-IR dark galaxies have been traced by radio emission and the detection by \champs~provides a measurement of their far-IR luminosities, thus dust-obscured SFRs, and more precise redshift estimates.

Figure~\ref{fig:mirimap} shows the coverage of a large area ($13\arcmin\times7\arcmin$, centered on PRIMER-COSMOS) by JWST/MIRI $7.7\,{\rm \mu m}$ \citep[left, from COSMOS-Web;][]{franco25}, MeerKAT $1\,{\rm GHz}$ \citep[middle, from MIGHTEE;][]{Jarvis2016,Hale2025}, and \champs~(right). \champs~detections at $>5\sigma$ are indicated by circles, labeled with the optical photometric redshifts provided by the COSMOS2025 catalog \citep{shuntov25}.
The right panels show \texttt{CIGALE} fits including the COSMOS20205 and \champs~photometry, for two example galaxies at $z\approx5.6$ and $z\approx5.3$, exemplifying the dusty nature of these sources. Note that without the \champs~data point, the far-IR, hence their dust mass and obscured star formation, would be largely unconstrained (indicated by the dashed lines).

%%% FIGURE: EMISSION LINES  %%%%
\begin{figure*}[t!]
\centering
\includegraphics[angle=0,width=1.04\columnwidth]{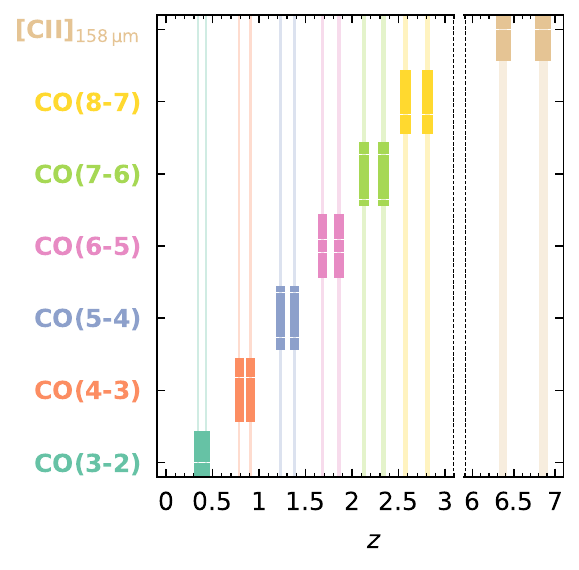}
\includegraphics[angle=0,width=1.06\columnwidth]{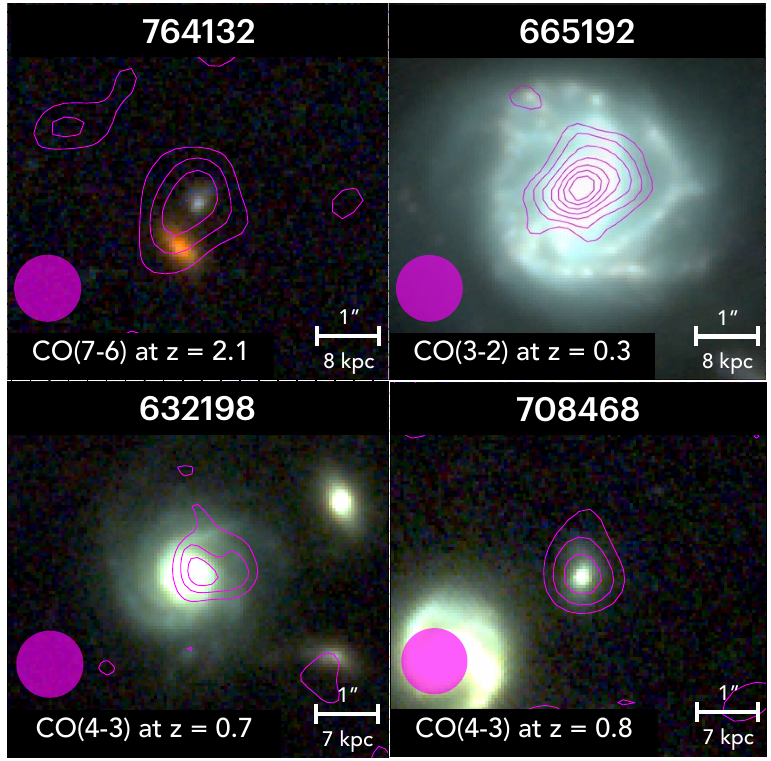}\vspace{-2mm}
\caption{
{\em Left:} Covered CO transitions as well as \Cii$_{\rm 158\mu m}$ emission line by the \champs~spectroscopic blind survey. The two vertical bands for each of the emission lines denote the two spectral windows in the side bands.
{\em Right:} Examples of CO-detected galaxies in the \champs~spectroscopic blind survey. The COSMOS2025 IDs, scale, and beam sizes are indicated. The CO contours ($3-9\sigma$ in steps of $1\sigma$) are shown in lilac on top of the JWST/NIRCam RGB images.
\label{fig:linedetection}
}%\vspace{-3mm}
\end{figure*}
%%%%%%%%%%%%%%%%%%%%%%%%%%%%%%%

\subsubsection{The Power of Stacking Across the Main-Sequence}\label{sec:sciencegoal3}

The large area coverage of \champs~provides the unique opportunity for stacking analyses. According to recent number counts on the COSMOS-Web field, there are more than 50\,000 MIRI detections and more than 245\,000 NIRCam detections\footnote{This assumes ${\rm SNR}>5$ detections in NIRCam and MIRI.} \citep{shuntov25} over the \champs~area.
The availability of this many sources allows for a detailed stacking analysis in fine bins of redshift and galaxy properties (such as stellar mass, SFR, or UV reddening). Classical binning, or more sophisticated unsupervised machine learning methods \citep[e.g.,][]{Lin2026}, can be applied to obtain stacked \champs~photometry and comprehensive SEDs. 
This opens up several new avenues of science investigations such as
{\em (i)} a comprehensive study of ISM mass fractions on and off the main sequence \citep[similar to lower-redshift studies with Herschel;][]{Santini2014},
{\em (ii)} deriving limits on the (currently undetected) far-IR emission of the elusive ``little red dots`` \citep[LRDs;][]{Greene2024,Matthee2024,Kocevski2025,Kocevski2023}, and 
{\em (iii)} constraining the average dust masses (or dust mass limits) of quiescent galaxies at $z>3$.

The potential of stacking hundreds to thousands of sources per bin will significantly lower the infrared luminosity survey limit by an order of magnitude to $10^{10}-10^{11}\,{\rm L_\odot}$ (Figure~\ref{fig:sensitivity}, left), equivalent to a multi-hour integration per pointing. 

Stacked \champs~data has already been used.
In \citet{Akins2025}, $\sim30\%$ of the \champs~early data of 434 LRDs at $z>5$ were stacked to put constraints on their far-IR emission. Intriguingly, the LRDs lack sub-millimeter detections (even in stacks of hundreds), which is curious given their substantial rest-frame optical SED-derived dust attenuation ($A_{\rm V}>2\,{\rm mag}$). This is consistent with independent non-detections from ALMA stacking of $60$ LRDs \citep{Casey2025}, and ALMA$+$MIRI imaging of two luminous LRDs \citep{Setton2025}.
This could suggest their continuum emission is dominated by non-stellar contributions (as this high $A_{\rm V}$ would imply more IR emission due to energy-balance arguments) as discussed in \citet{Akins2025}. However, as mentioned previously, far-IR luminosity is highly degenerate with dust temperature, and the non-detections of far-IR emission could be explained by a larger fraction of hot dust close to the AGN itself.

\citet{Zavala2026} has used $400$ \champs~detections to inform a new selection method for faint dusty star forming galaxies out to $z\approx 8$. The selection methodology employs a combined criterion based on stellar mass, star formation properties, and JWST/NIRCam color diagnostics. This improved method reveals a class of intrinsically faint, dust-enshrouded galaxies that have been mostly missed by earlier (sub)millimeter surveys, as well as by previous observations from both ground-based and space-based facilities. The results indicate that this population may represent a crucial evolutionary link between UV-luminous galaxies at $z>10$ and massive quiescent systems observed at $z<5$.

More recently, \citet{casey2026} mapped the average dust attenuation, emission, and opacity across cosmic time by stacking hundreds of galaxies for a detection in various sub-mm bands including \champs~data.

\subsection{A Blind-Survey of CO and \Cii~Line Emission}\label{sec:sciencegoal4}

In addition to continuum imaging, \champs~also provides a free blind-survey of strong far-IR line emitters at $1.2\,{\rm mm}$ at a $5\sigma$ peak line sensitivity of $6\,{\rm mJy}$ (over one-third of a line with a width of $250\,{\rm km\,s^{-1}}$).
As shown in the left panel of Figure~\ref{fig:linedetection}, \champs~covers CO(3-2), CO(4-3), CO(5-4), CO(6-5), CO(7-6), and CO(8-7) at approximate redshifts of $0.4$, $0.8$, $1.3$, $1.7$, $2.2$, and $2.7$. The spectral reductions are still ongoing and will be presented in an upcoming paper.

Based on the CO number densities observed by ASPECS \citep{Boogaard2020,Decarli2020}, \champs~is expected to detect approximately $20-110$ CO line emitters out to CO(8-7).
Assuming the CO SLED from \citet{Boogaard2021}, \champs~is sensitive to molecular gas estimates around $2\times10^9\,{\rm M_\odot}$ at $z = 2.2$ from CO(7-6).
The right panel of Figure~\ref{fig:linedetection} show example detections of CO emission in four galaxies at different redshifts.

\champs~has also the potential to put constraints on the bright-end of the \Cii$_{158\,{\rm \mu m}}$ luminosity function at $>10^9\,{\rm L_\odot}$ at $z\sim6.4$ and $z\sim6.8$. However, given the narrow frequency range and shallow depth, no more than $\approx5$ individual \Cii$_{158{\rm \mu m}}$~detections are expected assuming the models by \citet{Lagache2018} or \citet{Bethermin2022} (but see, e.g., \citet{liulunjun24} or \citet{lunjunliu26} for a more recent model of \Cii$_{158{\rm \mu m}}$~in high-$z$ galaxies that agrees better with observations and may suggest a higher number count of bright \Cii$_{158{\rm \mu m}}$~emitters).
On the other hand, \champs~can place stringent constraints on the \Cii$_{158{\rm \mu m}}$~emission of $z\sim6.6$ narrow-band-selected \lya~emitters \citep[e.g.,][]{Ouchi2018} through a systematic stacking analysis.

\section{Conclusions}\label{sec:summary}

In this paper, we presented \champs, a new ALMA $1.2\,{\rm mm}$ (band) blind-field survey covering $0.2\,{\rm deg^2}$ of the COSMOS-Web field. \champs~is aligned with both JWST/NIRCam and MIRI photometric observations, offering new multi-wavelength studies.
The uniform synthesized beam size of the full mosaic is $1.1\arcsec$ and the $1\sigma$ median RMS is $135\,{\rm \mu Jy/beam}$. This corresponds to a $5\sigma$ total IR luminosity per beam between $3\times10^{11}\,{\rm L_\odot}$ ($T_{\rm dust} = 30\,{\rm K}$) and $2\times10^{12}\,{\rm L_\odot}$ ($T_{\rm dust} = 60\,{\rm K}$) at $z=6$.
Compared to other sub-mm surveys, \champs~covers a unique niche in terms of large area and sensitivity.
For details on the data reduction and source extraction, we refer to \citet{Martinez2026}.

\champs's large area reveals hundreds of extremely dust-obscured sources, including more than a dozen near-IR dark galaxies at $z>6$. Contrary to current surveys tied to targeted ALMA large programs, \champs~provides a clean selection function and unbiased view on the most rare dusty sources.
The $1.2\,{\rm mm}$ observations by \champs~provide the crucial anchor to ancillary coverage at X-ray (Chandra), optical to mid-IR (Hubble and JWST), $2\,{\rm mm}$ (ALMA), and radio (MeerKAT) wavelengths to improve redshifts and derive robust physical parameters of the most dusty galaxies in the universe.

The main science goals include the discovery of the most dusty galaxies and AGN in the early universe, the characterization of their full SEDs, and population studies on and off the star-forming main-sequence using stacking of hundreds to thousands of galaxies and AGN.

As demonstrated this paper, \champs~has already identified three heavily dust-obscured high-redshift galaxy candidates based on $1.2\,{\rm mm}$ vs. JWST near-IR flux ratios. These galaxies, likely living in the EoR, belong to some of the most dust obscured (undetected in JWST/F150W and two even in JWST/F277W) and most massive ($\rm \log(M/M_\odot) > 10.4$) galaxies ever detected.

Furthermore, \champs~enables a spectroscopic blind-survey of CO lines between $z\approx0.4$ and $z\approx2.7$ as well as \Cii$_{\rm 158\mu m}$ at $z\approx6.7$.

Finally, the large parameter space covered by \champs~provides various opportunities for follow-up for an even more detailed look at the most dusty sources in the early universe to learn about their growth and dust production mechanisms.

%% Please use the acknowledgment and contribution environments. This will 
%% be anonomyized when the "anonymous" style option is used. 
\begin{acknowledgments}
{\em Acknowledgments:} 
The authors are grateful to George Privon and the NRAO computing team for their assistance with running the data reduction and portions of the data analysis on the NRAO-designated computing resources.
JK and FM also thank NRAO for hosting them during the initial phases of the data reduction and for providing valuable guidance on the data reduction procedures. FM acknowledges
support for this work that was provided by the NSF through award SOSP 1519126 from the NRAO.
The National Radio Astronomy Observatory and Green Bank Observatory are facilities of the U.S.
National Science Foundation operated under cooperative agreement by Associated Universities, Inc.
This paper makes use of the following ALMA data: ADS/JAO.ALMA\#2023.1.00180.L. ALMA is a
partnership of ESO (representing its member states), NSF (USA) and NINS (Japan), together with NRC
(Canada), MOST and ASIAA (Taiwan), and KASI (Republic of Korea), in cooperation with the Republic of
Chile. The Joint ALMA Observatory is operated by ESO, AUI/NRAO and NAOJ.
ET acknowledges support from the ANID CATA-BASAL program FB210003, and FONDECYT Regular 1241005 and 1250821.
HSBA gratefully acknowledges support from Academia Sinica through grant AS-PD-1141-M01-2.
DBS gratefully acknowledges support from NSF Grant 2407752.
The Cosmic Dawn Center (DAWN) is funded by the Danish National Research Foundation (DNRF140).
This work made use of \texttt{OverCite} \citep{Shariat2026}, an in-editor citation tool for \LaTeX.
\end{acknowledgments}

\begin{contribution}
%%This section gives authors the space to recognize author contributions. The text inside this environment is NOT counted towards the total word quanta. At a minimum, manuscripts are expected to include this text:

ALF is the PI of the \champs~ALMA large program and responsible for writing and submitting this manuscript. MA, CC, JK, JS, ST, ET, and JZ are co-PIs of \champs~and significantly contributed to the survey. FM led the reduction of the data and contributed significantly to this manuscript. All other authors contributed by providing comments to this manuscript or were part of the original \champs~ALMA proposal.
%All authors contributed equally to the \champs~collaboration.

%% But authors are expected to provide more specific details, e.g. 
%%
%%SC was responsible for writing and submitting the manuscript.
%%WWM came up with the initial research concept and edited the manuscript.
%%OTS obtained the funding and edited the manuscript.
%%EBF provided the formal analysis and validation. He also edited the manuscript.
%%GEH Supervised the undergraduates, wrote the software and administers the project github and Zenodo repositories.
%%
%% Authors can use the Contributor Role Taxonomy (CRediT) at
%% for ideas on how write a good statement tailored to their needs.

\end{contribution}

\facilities{ALMA}

\software{astropy \citep{astropy13,astropy18,astropy22}, 
CASA \citep{Hunter2023}
          }


%\newpage
%\appendix
%\section{Cutouts}

%% For this sample we use BibTeX plus aasjournalv7.bst to generate the

\bibliography{bibli}{}
\bibliographystyle{aasjournalv7}



\allauthors
\end{document}